\documentclass[final,3p,times]{elsarticle}
\usepackage{amssymb}
\usepackage[fleqn]{amsmath}
\usepackage{amsthm}
\usepackage{subfigure}
\usepackage{graphicx}

\biboptions{numbers,sort&compress}
\usepackage{hyperref}

\journal{}

\begin{document}

\begin{frontmatter}



\title{Generalized Darboux transformation and localized waves  in coupled Hirota equations }

\author{Xin Wang} \author{Yuqi Li}
\author{Yong Chen\corref{cor1}}
\ead{ychen@sei.ecnu.edu.cn}

\cortext[cor1]{Corresponding author.}

\address{Shanghai Key Laboratory of Trustworthy Computing, East China Normal University, Shanghai, 200062, People's Republic of China}

\begin{abstract}
In this paper, we construct a generalized Darboux transformation to the coupled Hirota equations
with high-order nonlinear effects like the third dispersion, self-steepening and inelastic Raman scattering terms.
As application, an $N$th-order localized wave solution on the plane backgrounds with the same spectral parameter
is derived through the direct iterative rule.
In particular, some semi-rational, multi-parametric localized wave solutions are obtained:
(1) Vector generalization of the first- and the second-order rogue wave solution;
(2) Interactional solutions between a dark-bright soliton  and a rogue wave, two dark-bright solitons and a
second-order rogue wave;
(3) Interactional solutions between a breather and a rogue wave, two breathers and a second-order rogue wave.
The results further reveal the striking dynamic structures of localized  waves in complex coupled systems.
\end{abstract}

\end{frontmatter}



\section{Introduction}
In the past decades, localized  waves including  bright or dark solitons , breathers
and rogue waves have attracted widespread attention in the research field of mathematical physics \cite{ad01,ad02,ad03,ad04,az01,w1,w2,w3,1,2,3,6}.
A breather is localized in space or time, namely, Akhmediev breather \cite{ad03} or Ma breather \cite{ad04}. While
a rogue wave is localized in both space and time, it appears from nowhere and disappears without a trace \cite{6}, and
has become a hottest topic in the research field of localized waves in very recent years.
Many nonlinear single-component systems are found to possess rogue wave solution,
the nonlinear Schr\"{o}dinger (NLS) equation \cite{4,5,6,7,8,9,10,11,21,22}, the derivative NLS equation \cite{18,23},
the Hirota equation \cite{19}, the Sasa-Satsuma equation \cite{32},
the variable coefficient  NLS equation \cite{31}, the discrete NLS equation \cite{30},
the Davey-Stewartson (DS) equation \cite{37}, etc.
Nevertheless, as is known to us,
some complex systems such as Bose-Einstein condensates, nonlinear optical fibers always involve more than a single component \cite{27,29,exx01},
so the latest studies on rogue waves or the other kinds of localized waves
have gradually been focused on the multi-component systems, in which the localized  waves
can present many novel peculiar phenomenons such as dark rogue waves in the coupled Gross-Pitaevskii equations \cite{26},
four-petaled flower rogue waves in the three-component NLS equations \cite{29}.
Notably, in coupled systems,
interactions between  Peregrine soliton and the other
nonlinear waves have become a hot topic of great interest, which has been explicitly shown that a Peregrine soliton
attracts a dark-bright soliton or a breather in the Manakov system \cite{24,27,ex02,mu}.

It is well known that the classical Darboux transformation (DT) \cite{16,17,dt1,dt2,dt3,dt4} can be iterated one by one such that
all spectrum parameters are chosen differently, or there will be singularities in the elements of Darboux matrix.
Consequently, the classical DT can not be directly used to construct rogue wave solutions or
the more complicated localized wave solutions of the nonlinear equations.
To overcome this problem, the so-called generalized DT was put forward by Guo, Ling and Liu to
investigate rogue wave solutions of the scalar NLS equation \cite{22},
then it was soon successfully applied to the Hirota equation \cite{22},
the derivative NLS equation \cite{23}, the AB model \cite{AB} and so on \cite{ex01}.
What's more, recently, Ling, Guo and Zhao generalize this method to research  higher-order
rogue wave solutions of the vector NLS equations \cite{33} (Manakov model) with the aid of the
corresponding  $3\times3$ matrix spectra. They find some striking phenomenons that four or six fundamental rogue waves
can emerge for second-order vector rogue wave in the coupled system.
Hence the generalized DT also provides an effective way to find higher-order localized  wave solutions of the
coupled systems.

In this paper, motivated by the work of Baronio \cite{27} and Guo \cite{22,33},
we discuss localized  wave solutions of the coupled Hirota (CH) equations
\begin{align}
& \mathrm{i}u_{t}+\frac{1}{2}u_{xx}+(|u|^{2}+|v|^{2})u+\mathrm{i}\epsilon[u_{xxx}+(6|u|^{2}+3|v|^{2})u_{x}+3uv^{*}v_{x}]=0,  \label{01} \\
& \mathrm{i}v_{t}+\frac{1}{2}v_{xx}+(|u|^{2}+|v|^{2})v+\mathrm{i}\epsilon[v_{xxx}+(6|v|^{2}+3|u|^{2})v_{x}+3vu^{*}u_{x}]=0.  \label{02}
\end{align}
Here $u(x,t)$, $v(x,t)$ are the complex smooth envelop functions, and $\epsilon$ is a small dimensionless real parameter.
The CH equations with high-order effects like the third dispersion, self-steepening and inelastic Raman scattering terms
were first proposed by Tasgal and Potasek to describe a non-relativistic boson field \cite{39}. They are important in
optics to illustrate the transmission when pulse lengths become comparable to the wavelength, while in this case the simple
Manakov model is inadequate, and the high-order nonlinear effects must be considered \cite{41}.
Some important results have been obtained for Eqs. (1) and (2),
such as the Lax pair, the classical  Darboux transformation, the Painlev\'{e} analysis,
the bright and dark soliton solutions \cite{39,40,41}.
Especially, in Ref. \cite{28}, Chen and Song give the lower-order
fundamental and dark rogue wave solutions of Eqs. (1) and (2),  which extremely indicate the interesting structures of
localized waves in Eqs. (1) and (2).

In the present paper, we concentrate on interactional solutions between rogue waves and the other nonlinear waves such as
dark-bright solitons and breathers of Eqs. (1) and (2), which, to the best of our knowledge have not been reported by any authors.
By resorting to the Taylor series expansion coefficients of a special solution to the linear spectral problem
of Eqs. (1) and (2),  a generalized DT with several free parameters is constructed. As application,
a unified formula of $N$th-order localized wave solution on the plane backgrounds with the same spectral parameter
is derived through the direct iterative rule. In particular,
apart from the vector generalization of the first- and the second-order rogue wave solutions
to the decoupled Hirota equation,
some novel localized wave solutions of Eqs. (1) and (2) are provided, such as
interactional solutions between a dark-bright soliton and a rogue wave, two dark-bright solitons and a second-order
rogue wave, and interactional solutions between a breather and a rogue wave, two breathers and a second-order
rogue wave. The free parameters play a crucial role to affect the dynamic distributions of
localized waves in the coupled system. Some different types of figures by adjusting the free parameters
are explicitly shown to illustrate the dynamic properties of the localized nonlinear waves. Our results can be
seen as the generalization of the work reported by Baronio et al. \cite{27} to the complex coupled system
with high-order nonlinear terms.

The paper is organized as follows. In section 2, the generalized DT of Eqs. (1) and (2) is constructed.
In section 3, some explicit general localized wave solution and
interesting figures are given.  The last section contains some discussion.

\section{Generalized Darboux transformation}
In this section, we construct the generalized DT to Eqs. (1) and (2). The Lax pair of it can be expressed as \cite{39}
\begin{align}
&\Phi_{x}=U\Phi,\ \ U=\zeta U_{0}+U_{1}, \label{03} \\
&\Phi_{t}=V\Phi,\ \ V=\zeta^{3}V_{0}+\zeta^{2}V_{1}+\zeta V_{2}+V_{3},\label{04}
\end{align}
where
$$
U_{0}=\frac{1}{12\epsilon}\left(
\begin{array}{ccc}
-2\mathrm{i} & 0 & 0 \\
0 & \mathrm{i} & 0 \\
0 & 0 & \mathrm{i} \\
\end{array}
\right),\ \
U_{1}=
\left(
  \begin{array}{ccc}
    0 & -u & -v \\
    u^{*} & 0 & 0 \\
    v^{*} & 0 & 0 \\
  \end{array}
\right),\ V_{0}=\frac{1}{16\epsilon}U_{0},V_{1}=\frac{1}{8\epsilon}U_{0}+\frac{1}{16\epsilon}U_{1},
$$
$$
V_{2}=\frac{1}{4}
\left(
  \begin{array}{ccc}
    \mathrm{i}e & -\frac{u}{2\epsilon}-\mathrm{i}u_{x} &  -\frac{v}{2\epsilon}-\mathrm{i}v_{x}\\
    \frac{u^{*}}{2\epsilon}-\mathrm{i}u^{*}_{x} & -\mathrm{i}|u|^{2} & -\mathrm{i}vu^{*} \\
    \frac{v^{*}}{2\epsilon}-\mathrm{i}v^{*}_{x} & -\mathrm{i}uv^{*} & -\mathrm{i}|v|^{2} \\
  \end{array}
\right),\
V_{3}=
\left(
  \begin{array}{ccc}
    \epsilon(e_{1}+e_{2})+\frac{\mathrm{i}}{2}e & \epsilon e_{3}-\frac{\mathrm{i}}{2}u_{x} & \epsilon e_{4}-\frac{\mathrm{i}}{2}v_{x} \\
    -\epsilon e_{3}^{*}-\frac{\mathrm{i}}{2}u^{*}_{x} & -\epsilon e_{1}-\frac{\mathrm{i}}{2}|u|^{2} & \epsilon e_{5}-\frac{\mathrm{i}}{2}vu^{*} \\
    -\epsilon e_{4}^{*}-\frac{\mathrm{i}}{2}v^{*}_{x} & -\epsilon e^{*}_{5}-\frac{\mathrm{i}}{2}uv^{*} & -\epsilon e_{2}-\frac{\mathrm{i}}{2}|v|^{2} \\
  \end{array}
\right),
$$
with
$$
e=|u|^{2}+|v|^2,\  e_{1}=uu^{*}_{x}-u^{*}u_{x},\ e_{2}=vv^{*}_{x}-v^{*}v_{x},\ e_{3}=u_{xx}+2eu,\ e_{4}=v_{xx}+2ev,\ e_{5}=u^{*}v_{x}-vu^{*}_{x}.
$$
Here $\Phi=(\psi(x,t),\phi(x,t),\chi(x,t))^{T}$, $u$ and $v$ are potentials, $\zeta$ is  the spectral parameter, $u^{*}$ and $v^{*}$ denote
complex conjugate of $u$ and $v$. Through direct calculation, one can directly get Eqs. (1) and (2) by
using the zero curvature equation $U_{t}-V_{x}+UV-VU=0$.

After that,  let $\Phi_{1}=(\psi_{1},\phi_{1},\chi_{1})^{T}$ be a solution of (\ref{03}) and (\ref{04}) with $u=u[0]$, $v=v[0]$ and  $\zeta=\zeta_{1}$, then
based on the classical DT  of the Ablowitz-Kaup-Newell-Segur (AKNS) spectral problem \cite{16}, the following formulas
\begin{align}
&\Phi[1]=T[1]\Phi,\ T[1]=\zeta I-H[0]\Lambda_{1}H[0]^{-1},\label{05}\\
&u[1]=u[0]+{\rm i}(\zeta_{1}-\zeta_{1}^{*})\frac{\psi_{1}[0]\phi_{1}[0]^{*}}{4\epsilon(|\psi_{1}[0]|^{2}+|\phi_{1}[0]|^{2}+|\chi_{1}[0]|^{2})},\label{06}\\
&v[1]=v[0]+{\rm i}(\zeta_{1}-\zeta_{1}^{*})\frac{\psi_{1}[0]\chi_{1}[0]^{*}}{4\epsilon(|\psi_{1}[0]|^{2}+|\phi_{1}[0]|^{2}+|\chi_{1}[0]|^{2})},\label{07}
\end{align}
satisfy
\begin{equation}\label{08}
\Phi[1]_{x}=U[1]\Phi[1],\ \Phi[1]_{t}=V[1]\Phi[1],
\end{equation}
where $\psi_{1}[0]=\psi_{1}$, $\phi_{1}[0]=\phi_{1}$, $\chi_{1}[0]=\chi_{1}$,
$$
I=\left(
\begin{array}{ccc}
1 & 0 & 0\\
0 & 1 & 0\\
0 & 0 & 1
\end{array}
\right),\ \
H[0]=\left(
\begin{array}{ccc}
\psi_{1}[0] & \phi_{1}[0]^{*} & \chi_{1}[0]^{*}\\
\phi_{1}[0] & -\psi_{1}[0]^{*} &  0\\
\chi_{1}[0] &  0 & -\psi_{1}[0]^{*}
\end{array}
\right),\ \
\Lambda_{1}=\left(
 \begin{array}{ccc}
 \zeta_{1} & 0 & 0\\
 0 & \zeta_{1}^{*} & 0 \\
 0 & 0 & \zeta_{1}^{*}
 \end{array}
 \right),
$$
$U[1],~V[1]$ has the same form as $U,~V$ except the old potentials $u$, $v$ are replaced by the new ones $u[1]$, $v[1]$, and (\ref{05})-(\ref{07}) are called
the classical DT of the Lax pair (\ref{03}) and (\ref{04}).

Thus, assume $\Phi_{l}=(\psi_{l},\phi_{l},\chi_{l})^{T} (1\leq l\leq N)$ be a basic solution of the
Lax pair (\ref{03}) and (\ref{04}) with $u=u[0]$, $v=v[0]$ and $\zeta=\zeta_{l}$, then
by making use of  the above DT $N$ times, we get the $N$-step DT of Eqs. (1) and (2)
\begin{align}
&\Phi[N]=T[N]T[N-1]\cdots T[1]\Phi, \ T[l]=\zeta I-H[l-1]\Lambda_{l}H[l-1]^{-1},\label{09}\\
&u[N]=u[N-1]+{\rm i}(\zeta_{N}-\zeta_{N}^{*})\frac{\psi_{N}[N-1]\phi_{N}[N-1]^{*}}{4\epsilon(|\psi_{N}[N-1]|^{2}+|\phi_{N}[N-1]|^{2}
+|\chi_{N}[N-1]|^{2})},\label{10}\\
&v[N]=v[N-1]+{\rm i}(\zeta_{N}-\zeta_{N}^{*})\frac{\psi_{N}[N-1]\chi_{N}[N-1]^{*}}{4\epsilon(|\psi_{N}[N-1]|^{2}+|\phi_{N}[N-1]|^{2}
+|\chi_{N}[N-1]|^{2})},\label{11}
\end{align}
where
$$
H[l-1]=\left(
\begin{array}{ccc}
\psi_{l}[l-1] & \phi_{l}[l-1]^{*} & \chi_{l}[l-1]^{*}\\
\phi_{l}[l-1] & -\psi_{l}[l-1]^{*} &  0\\
\chi_{l}[l-1] &  0 & -\psi_{l}[l-1]^{*}
\end{array}
\right),\ \Lambda_{l}=\left(
 \begin{array}{ccc}
 \zeta_{l} & 0 & 0\\
 0 & \zeta_{l}^{*} & 0 \\
 0 & 0 & \zeta_{l}^{*}
 \end{array}
 \right),
$$
with $(\psi_{l}[l-1],\phi_{l}[l-1],\chi_{l}[l-1])^{T}=\Phi_{l}[l-1]$ and
$$
\Phi_{l}[l-1]=T_{l}[l-1]T_{l}[l-2]\cdots T_{l}[1]\Phi_{l},\  T_{l}[j]=T[j]|_{\zeta=\zeta_{l}},\ 1\leq l\leq N,~1\leq j\leq l-1.
$$

In the following, according to the above classical DT (\ref{05})-(\ref{07}) , we construct
the generalized DT to Eqs. (1) and (2). Denote
\begin{equation}\label{12}
\Phi_{1}=\Phi_{1}(\zeta_{1}+\delta)
\end{equation}
is a special solution of (\ref{03}) and (\ref{04}) with $u=u[0]$, $v=v[0]$ and $\zeta=\zeta_{1}+\delta$. Here
$\Phi_{1}=(\psi_{1},\phi_{1},\chi_{1})^{T}$, $\delta$ is a small
parameter. Next it is  significant that we suppose  $\Phi_{1}$ be expanded as the Taylor series
\begin{equation}\label{13}
\Phi_{1}=\Phi_{1}^{[0]}+\Phi_{1}^{[1]}\delta+\Phi_{1}^{[2]}\delta^{2}+\Phi_{1}^{[3]}\delta^{3}+\cdots+\Phi_{1}^{[N]}\delta^{N}+\cdots,
\end{equation}
where $\Phi_{1}^{[l]}=(\psi_{1}^{[l]},\phi_{1}^{[l]},\chi_{1}^{[l]})$, $\Phi_{1}^{[l]}=\frac{\displaystyle1}{\displaystyle l!}
\frac{\displaystyle\partial^{l}\Phi_{1}}{\displaystyle\partial \delta^{l}}\mid_{\delta=0} (l=0,1,2,\cdots).$

From the above assumption,
it is easy to find that $\Phi_{1}^{[0]}$ is a particular solution of  (\ref{03}) and (\ref{04}) with  $u=u[0]$, $v=v[0]$ and $\zeta=\zeta_{1}$.
So the first-step generalized DT can be directly given by means of the formulas (\ref{05})-(\ref{07}).

(1) The first-step generalized DT.
\begin{align}
&\Phi[1]=T[1]\Phi,\ T[1]=(\zeta I-H[0]\Lambda_{1}H[0]^{-1}),\label{14}\\
&u[1]=u[0]+{\rm i}(\zeta_{1}-\zeta_{1}^{*})\frac{\psi_{1}[0]\phi_{1}[0]^{*}}{4\epsilon(|\psi_{1}[0]|^{2}+|\phi_{1}[0]|^{2}+|\chi_{1}[0]|^{2})},\label{15}\\
&v[1]=v[0]+{\rm i}(\zeta_{1}-\zeta_{1}^{*})\frac{\psi_{1}[0]\chi_{1}[0]^{*}}{4\epsilon(|\psi_{1}[0]|^{2}+|\phi_{1}[0]|^{2}+|\chi_{1}[0]|^{2})},\label{16}
\end{align}
where $\psi_{1}[0]=\psi_{1}^{[0]},~\phi_{1}[0]=\phi_{1}^{[0]},$~$\chi_{1}[0]=\chi_{1}^{[0]}$ and
$$
H[0]=\left(
\begin{array}{ccc}
\psi_{1}[0] & \phi_{1}[0]^{*} & \chi_{1}[0]^{*}\\
\phi_{1}[0] & -\psi_{1}[0]^{*} &  0\\
\chi_{1}[0] &  0 & -\psi_{1}[0]^{*}
\end{array}
\right),\ \
\Lambda_{1}=\left(
 \begin{array}{ccc}
 \zeta_{1} & 0 & 0\\
 0 & \zeta_{1}^{*} & 0 \\
 0 & 0 & \zeta_{1}^{*}
 \end{array}
 \right).
$$

(2) The second-step generalized DT.

It is obvious that $T[1]|\Phi_{1}$ is a solution of the Lax pair  (\ref{03}) and (\ref{04}) with $u[1]$, $v[1]$,
$\zeta=\zeta_{1}+\delta$,
and so is $T[1]\Phi_{1}/\delta$. Consequently,  considering the following limit process

$$\begin{array}{l}
\lim\limits_{\delta\rightarrow0}\frac{\displaystyle T[1]|_{\zeta=\zeta_{1}+\delta}\Phi_{1}}{\displaystyle\delta}=\lim\limits_{\delta\rightarrow0}
\frac{\displaystyle(\delta+T_{1}[1])\Phi_{1}}{\displaystyle\delta}=
\Phi_{1}^{[0]}+T_{1}[1]\Phi_{1}^{[1]}\equiv\Phi_{1}[1],
\end{array}$$
we find a nontrivial solution of the Lax pair  (\ref{03}) and (\ref{04}) with $u[1]$, $v[1]$, $\zeta=\zeta_{1}$, and here we have used the identity
$T_{1}[1]\Phi_{1}^{[0]}=0.$ Then the second-step generalized DT holds
\begin{align}
&\Phi[2]=T[2]T[1]\Phi,\ T[2]=(\zeta I-H[1]\Lambda_{2}H[1]^{-1}),\label{17}\\
&u[2]=u[1]+{\rm i}(\zeta_{1}-\zeta_{1}^{*})\frac{\psi_{1}[1]\phi_{1}[1]^{*}}{4\epsilon(|\psi_{1}[1]|^{2}+|\phi_{1}[1]|^{2}+|\chi_{1}[1]|^{2})},\label{18}\\
&v[2]=v[1]+{\rm i}(\zeta_{1}-\zeta_{1}^{*})\frac{\psi_{1}[1]\chi_{1}[1]^{*}}{4\epsilon(|\psi_{1}[1]|^{2}+|\phi_{1}[1]|^{2}+|\chi_{1}[1]|^{2})},\label{19}
\end{align}
where $(\psi_{1}[1],\phi_{1}[1],\chi_{1}[1])^{T}=\Phi_{1}[1],$
$$
H[1]=\left(
\begin{array}{ccc}
\psi_{1}[1] & \phi_{1}[1]^{*} & \chi_{1}[1]^{*}\\
\phi_{1}[1] & -\psi_{1}[1]^{*} &  0\\
\chi_{1}[1] &  0 & -\psi_{1}[1]^{*}
\end{array}
\right),\ \
\Lambda_{2}=\left(
 \begin{array}{ccc}
 \zeta_{1} & 0 & 0\\
 0 & \zeta_{1}^{*} & 0 \\
 0 & 0 & \zeta_{1}^{*}
 \end{array}
 \right).
$$

(3) The third-step generalized DT.

Similarly, by using the following limit
$$\lim\limits_{\delta\rightarrow0}\frac{\displaystyle [T[2]T[1]]|_{\zeta=\zeta_{1}+\delta}\Phi_{1}}{\displaystyle\delta^{2}}=
\lim\limits_{\delta\rightarrow0}\frac{\displaystyle (\delta+T_{1}[2])(\delta+T_{1}[1])\Phi_{1}}{\displaystyle\delta^{2}}=
\Phi_{1}^{[0]}+(T_{1}[2]+T_{1}[1])\Phi_{1}^{[1]}+T_{1}[2]T_{1}[1]\Phi_{1}^{[2]}\equiv\Phi_{1}[2],
$$
a special solution of the Lax pair (\ref{03}) and (\ref{04}) with $u[2]$, $v[2]$, $\zeta=\zeta_{1}$ can be obtained, and
the identities
$$T_{1}[1]\Phi_{1}^{[0]}=0,\ T_{1}[2](\Phi_{1}^{[0]}+T_{1}[1]\Phi_{1}^{[1]})=0$$
have been applied in the above derivation process. Then the third-step generalized DT can be given
\begin{align}
&\Phi[3]=T[3]T[2]T[1]\Phi,\ T[3]=(\zeta I-H[2]\Lambda_{3}H[2]^{-1}),\label{20}\\
&u[3]=u[2]+{\rm i}(\zeta_{1}-\zeta_{1}^{*})\frac{\psi_{1}[2]\phi_{1}[2]^{*}}{4\epsilon(|\psi_{1}[2]|^{2}+|\phi_{1}[2]|^{2}+|\chi_{1}[2]|^{2})},\label{21}\\
&v[3]=v[2]+{\rm i}(\zeta_{1}-\zeta_{1}^{*})\frac{\psi_{1}[2]\chi_{1}[2]^{*}}{4\epsilon(|\psi_{1}[2]|^{2}+|\phi_{1}[2]|^{2}+|\chi_{1}[2]|^{2})},\label{22}
\end{align}
where $(\psi_{1}[2],\phi_{1}[2],\chi_{1}[2])^{T}=\Phi_{1}[2]$,
$$
H[2]=\left(
\begin{array}{ccc}
\psi_{1}[2] & \phi_{1}[2]^{*} & \chi_{1}[2]^{*}\\
\phi_{1}[2] & -\psi_{1}[2]^{*} &  0\\
\chi_{1}[2] &  0 & -\psi_{1}[2]^{*}
\end{array}
\right),\ \
\Lambda_{3}=\left(
 \begin{array}{ccc}
 \zeta_{1} & 0 & 0\\
 0 & \zeta_{1}^{*} & 0 \\
 0 & 0 & \zeta_{1}^{*}
 \end{array}
 \right).
$$

(4) The $N$-step generalized DT.

Continuing the above process, the $N$-step generalized DT can be presented as follows
$$
\Phi_{1}[N-1]=\Phi_{1}^{[0]}+\sum_{l=1}^{N-1}T_{1}[l]\Phi_{1}^{[1]}+\sum_{l=1}^{N-1}\sum_{k=1}^{l-1}T_{1}[l]T_{1}[k]\Phi_{1}^{[2]}
+\cdots+T_{1}[N-1]T_{1}[N-2]
\cdots T_{1}[1]\Phi_{1}^{[N-1]},
$$
\begin{align}
&\Phi[N]=T[N]T[N-1]\cdots T[1]\Phi,\ T[l]=(\zeta I-H[l-1]\Lambda_{l}H[l-1]^{-1}),\label{23}\\
&u[N]=u[N-1]+{\rm i}(\zeta_{1}-\zeta_{1}^{*})\frac{\psi_{1}[N-1]\phi_{1}[N-1]^{*}}{4\epsilon(|\psi_{1}[N-1]|^{2}+|\phi_{1}[N-1]|^{2}
+|\chi_{1}[N-1]|^{2})},\label{24}\\
&v[N]=v[N-1]+{\rm i}(\zeta_{1}-\zeta_{1}^{*})\frac{\psi_{1}[N-1]\chi_{1}[N-1]^{*}}{4\epsilon(|\psi_{1}[N-1]|^{2}+|\phi_{1}[N-1]|^{2}
+|\chi_{1}[N-1]|^{2})},\label{25}
\end{align}
where $(\psi_{1}[N-1],\phi_{1}[N-1],\chi_{1}[N-1])^{T}=\Phi_{1}[N-1]$,
$$
H[l-1]=\left(
\begin{array}{ccc}
\psi_{1}[l-1] & \phi_{1}[l-1]^{*} & \chi_{1}[l-1]^{*}\\
\phi_{1}[l-1] & -\psi_{1}[l-1]^{*} &  0\\
\chi_{1}[l-1] &  0 & -\psi_{1}[l-1]^{*}
\end{array}
\right),\ \
\Lambda_{l}=\left(
 \begin{array}{ccc}
 \zeta_{1} & 0 & 0\\
 0 & \zeta_{1}^{*} & 0 \\
 0 & 0 & \zeta_{1}^{*}
 \end{array}
 \right),~1\leq l\leq N.
$$

What we should mention is that (\ref{24})-(\ref{25}) give rise to a unified formula of $N$th-order localized wave solution to
Eqs. (1) and (2) on the plane backgrounds with the same spectral parameter, and they can be
converted into the $3N\times 3N$ determinant representation by using the so-called Crum theorem \cite{17}.
However,  we prefer to using the iterative form of the Darboux transformation of degree one
rather than the high-order determinant representation, for it can be more conveniently calculated by the
computer. In the next section, we shall present some explicit localized wave solutions of Eqs. (1) and (2)
to illustrate how to use these formulas, some interesting figures are shown.

\section{Localized  wave solutions}  
In this section,  we start from a periodic seed solution of Eqs. (1) and (2),
\begin{equation}\label{26}
u[0]=d_{1}{\rm e}^{{\rm i}\theta},\ \ \ v[0]=d_{2}{\rm e}^{{\rm i}\theta}.
\end{equation}
Here $\theta=(d_{1}^{2}+d_{2}^{2})t$, $d_{1}$ and $d_{2}$ are real constants.
Then the basic solution of the Lax pair (\ref{03}) and (\ref{04}) with $u[0]$, $v[0]$ and $\zeta$ holds

\begin{equation}\label{27}
\Phi_{1}=
\left(
  \begin{array}{c}
    (C_{1}{\rm e}^{M_{1}+M_{2}}-C_{2}{\rm e}^{M_{1}-M_{2}}){\rm e}^{\frac{{\rm i}}{2}\theta} \\
    \rho_{1}(C_{2}{\rm e}^{M_{1}+M_{2}}-C_{1}{\rm e}^{M_{1}-M_{2}}){\rm e}^{-\frac{{\rm i}}{2}\theta}+d_{2}\alpha{\rm e}^{M_{3}} \\
    \rho_{2}(C_{2}{\rm e}^{M_{1}+M_{2}}-C_{1}{\rm e}^{M_{1}-M_{2}}){\rm e}^{-\frac{{\rm i}}{2}\theta}-d_{1}\alpha{\rm e}^{M_{3}} \\
  \end{array}
\right),
\end{equation}
where
$$
\begin{array}{l}
C_{1}=\frac{\displaystyle(\zeta-\sqrt{\zeta^{2}+64\epsilon^{2}(d_{1}^{2}+d_{2}^{2})})^{\frac{1}{2}}} {\displaystyle\sqrt{\zeta^{2}+64\epsilon^{2}(d_{1}^{2}+d_{2}^{2})}},~~
C_{2}=\frac{\displaystyle(\zeta+\sqrt{\zeta^{2}+64\epsilon^{2}(d_{1}^{2}+d_{2}^{2})})^{\frac{1}{2}}} {\displaystyle\sqrt{\zeta^{2}+64\epsilon^{2}(d_{1}^{2}+d_{2}^{2})}},\\
\rho_{1}=\frac{\displaystyle d_{1}}{\displaystyle\sqrt{d_{1}^{2}+d_{2}^{2}}},~~\rho_{2}=\frac{\displaystyle d_{2}}{\displaystyle\sqrt{d_{1}^{2}+d_{2}^{2}}},~~
M_{1}=-\displaystyle\frac{{\rm i}\zeta}{384\epsilon^{2}}(16\epsilon x+\zeta(\zeta+2)t),\
M_{3}=\displaystyle\frac{{\rm i}\zeta}{192\epsilon^{2}}(16\epsilon x+\zeta(\zeta+2)t),\\
M_{2}=\displaystyle\frac{{\rm i}}{128\epsilon^{2}}\sqrt{\zeta^{2}+64\epsilon^{2}(d_{1}^{2}+d_{2}^{2})}
(16\epsilon x+(\zeta(\zeta+2)-32\epsilon^{2}(d_{1}^{2}+d_{2}^{2}))t+\sum_{k=1}^{N}s_{k}f^{2k}).
\end{array}
$$
Here $f$ is a small parameter, $s_{k}=m_{k}+{\rm i}n_{k}$. $\alpha$, $m_{k}$, $n_{k}$  ($1\leq k\leq N$) are real constants.
Let $\zeta=8{\rm i}\epsilon\sqrt{d_{1}^{2}+d_{2}^{2}}(1+f^2)$,
and expanding the vector function $\Phi_{1}(f)$  at $f=0$, we have
\begin{equation}\label{28}
\Phi_{1}(f)=\Phi_{1}^{[0]}+\Phi_{1}^{[1]}f^{2}+\Phi_{1}^{[2]}f^{4}+\cdots,
\end{equation}
where
$$\begin{array}{l}
\psi_{1}^{[0]}=\displaystyle\frac{({\rm i}-1)}{4\sqrt{\epsilon}(d_{1}^{2}+d_{2}^{2})^{1/4}}(2\sqrt{d_{1}^{2}+d_{2}^{2}}
(x-6\epsilon d_{1}^{2}t-6\epsilon d_{2}^{2}t)+2{\rm i}d_{1}^{2}t+2{\rm i}d_{2}^{2}t+1){\rm e}^{\xi_{1}},\\
\phi_{1}^{[0]}=\displaystyle\frac{({\rm i}-1)d_{1}}{4\sqrt{\epsilon}(d_{1}^{2}+d_{2}^{2})^{3/4}}(2\sqrt{d_{1}^{2}+d_{2}^{2}}
(x-6\epsilon d_{1}^{2}t-6\epsilon d_{2}^{2}t)+2{\rm i}d_{1}^{2}t+2{\rm i}d_{2}^{2}t-1){\rm e}^{\xi_{2}}+d_{2}\alpha {\rm e}^{\xi_{3}},\\
\chi_{1}^{[0]}=\displaystyle\frac{({\rm i}-1)d_{2}}{4\sqrt{\epsilon}(d_{1}^{2}+d_{2}^{2})^{3/4}}(2\sqrt{d_{1}^{2}+d_{2}^{2}}
(x-6\epsilon d_{1}^{2}t-6\epsilon d_{2}^{2}t)+2{\rm i}d_{1}^{2}t+2{\rm i}d_{2}^{2}t-1){\rm e}^{\xi_{2}}-d_{1}\alpha {\rm e}^{\xi_{3}},\cdots
\end{array}
$$
with
$$\begin{array}{l}
\xi_{1}=\displaystyle\frac{1}{3}\sqrt{d_{1}^{2}+d_{2}^{2}}x+(d_{1}^{2}+d_{2}^{2})(\frac{5}{6}{\rm i}-\frac{4}{3}\epsilon\sqrt{d_{1}^{2}+d_{2}^{2}})t,\\
\xi_{2}=\displaystyle\frac{1}{3}\sqrt{d_{1}^{2}+d_{2}^{2}}x-(d_{1}^{2}+d_{2}^{2})(\frac{1}{6}{\rm i}+\frac{4}{3}\epsilon\sqrt{d_{1}^{2}+d_{2}^{2}})t,\\
\xi_{3}=-\displaystyle\frac{2}{3}\sqrt{d_{1}^{2}+d_{2}^{2}}x-(d_{1}^{2}+d_{2}^{2})(\frac{2}{3}{\rm i}-\frac{8}{3}\epsilon\sqrt{d_{1}^{2}+d_{2}^{2}})t.
\end{array}
$$

It is direct to verify  that $\Phi_{1}^{[0]}=(\psi_{1}^{[0]},\phi_{1}^{[0]},\chi_{1}^{[0]})^{T}$ is a special solution of the Lax pair
(\ref{03}) and (\ref{04}) with $u=u[0]$,
$v=v[0]$, and $\zeta=\zeta_{1}=8{\rm i}\epsilon\sqrt{d_{1}^{2}+d_{2}^{2}}$.
Hence,
by using the formulas (\ref{15}) and (\ref{16}), we arrive at
\begin{equation}\label{29}
u[1]=d_{1}{\rm e}^{{\rm i}\theta}+\displaystyle\frac{d_{1}{\rm e}^{{\rm i}\theta}(F_{1}+{\rm i}H_{1})+d_{2}G_{1}{\rm e}^{\eta_{1}}}{D_{1}
+K_{1}{\rm e}^{\eta_{2}}},\
v[1]=d_{2}{\rm e}^{{\rm i}\theta}+\displaystyle\frac{d_{2}{\rm e}^{{\rm i}\theta}(F_{1}+{\rm i}H_{1})-d_{1}G_{1}{\rm e}^{\eta_{1}}}{D_{1}
+K_{1}{\rm e}^{\eta_{2}}},
\end{equation}
where
$$\begin{array}{l}
F_{1}=-8(d_{1}^{2}+d_{2}^{2})x^{2}+96\epsilon (d_{1}^{2}+d_{2}^{2})^{2}xt-8(d_{1}^{2}+d_{2}^{2})^{2}(36\epsilon^{2}(d_{1}^{2}+d_{2}^{2})+1)t^{2}+2,\\
H_{1}=8(d_{1}^{2}+d_{2}^{2})t,\ D_{1}=4(d_{1}^{2}+d_{2}^{2})x^{2}-48\epsilon(d_{1}^{2}+d_{2}^{2})^{2}xt+4(d_{1}^{2}+d_{2}^{2})^{2}(36\epsilon^{2}(d_{1}^{2}
+d_{2}^{2})+1)t^{2}+1,\\
G_{1}=4(1-{\rm i})\sqrt{\epsilon}\alpha(d_{1}^{2}+d_{2}^{2})^{3/4}(2\sqrt{d_{1}^{2}+d_{2}^{2}}(x-6\epsilon d_{1}^{2}t
-6\epsilon d_{2}^{2}t)+2{\rm i}d_{1}^{2}t+2{\rm i}d_{2}^{2}t+1),\\
K_{1}=4\epsilon\alpha^{2}(d_{1}^{2}+d_{2}^{2})^{3/2},\ \eta_{1}=\displaystyle-\sqrt{d_{1}^{2}
+d_{2}^{2}}x+(d_{1}^{2}+d_{2}^{2})(\frac{3}{2}{\rm i}+4\epsilon\sqrt{d_{1}^{2}+d_{2}^{2}})t,\\
\eta_{2}=-2\sqrt{d_{1}^{2}+d_{2}^{2}}x+8\epsilon(d_{1}^{2}+d_{2}^{2})^{3/2}t.
\end{array}$$
It is straightforward to check that the above solution satisfies Eqs. (1) and (2) with the aid of Maple, and
in what follows, we discuss the dynamics of this solution through three different cases.

(i) $\alpha=0$. Then the vector localized wave solution (\ref{29}) is reduced to
\begin{equation}\label{30}
u[1]_{rw}=\displaystyle d_{1}{\rm e}^{{\rm i}\theta}(1+\frac{F_{1}+{\rm i}H_{1}}{D_{1}}),\
v[1]_{rw}=\displaystyle d_{2}{\rm e}^{{\rm i}\theta}(1+\frac{F_{1}+{\rm i}H_{1}}{D_{1}}),
\end{equation}
which is nothing but the vector generalization of the first-order rogue wave solution to the
decoupled Hirota equation.
Here, $u[1]_{rw}$ is merely proportional to $v[1]_{rw}$, and from the concrete expressions of them, we calculate that
the maximum amplitudes of $u[1]_{rw}$ and $v[1]_{rw}$ are three times more than their each average crest.

(ii) $\alpha\neq0$, $d_{1}\neq0$ and  $d_{2}=0$. In this case, the dark-bright-rogue wave solution can be generated.
Fig. 1 shows a dark-bright soliton together with a rogue wave, and Figs. 2 and 3
describe the explicit collision processes between a dark soliton and a rogue wave, a bright soliton and
a rogue wave, respectively. We see that in Figs. 2 and 3, a dark soliton and a bright soliton  are propagating
in the positive direction of $x$-axis, while at the time of $t=0$, a rogue wave suddenly appears from nowhere,
and these two different types of waves impact together with each other. Next, the rogue wave
soon disappears in the same abrupt way, and the solitons continue going ahead without any changing
of their amplitudes and velocities after the collision. The whole interactional process can be seen as elastic.
Moreover, by decreasing the value of $\alpha$, the dark-bright soliton and the rogue wave separate.
We notice that in Fig. 4(a), a dark soliton and a rogue wave emerge on the distribution of the
spacial-temporal structure, and the maximum amplitude of the rogue wave is 3. Nevertheless,
when the bright soliton and the rogue wave divide, it is shown that the
rogue wave can not be easily identified, see Fig. 4(b). The phenomenon can be easily
explained, for the amplitude of a rogue wave depend on that of its background wave, and
at this time the amplitude of the background wave in $v$ component is zero.

(iii) $\alpha\neq0$, $d_{1}\neq0$ and  $d_{2}\neq0$. At this point, the interactional solution between
an Akhmediev breather and a rogue wave can be given, see Figs. 5 and 6. We observe that by
increasing the value of $\alpha$, the Akhmediev breather and the rogue wave merge,
see Fig. 5.
While by decreasing the value of $\alpha$, the Akhmediev breather and the rogue wave separate,
see Fig. 6.

Next, considering the following limit
\begin{equation}\label{31}
\lim\limits_{f\rightarrow0}\frac{\displaystyle T[1]|_{\zeta=8{\rm i}\epsilon\sqrt{d_{1}^{2}+d_{2}^{2}}
(1+f^{2})}\Phi_{1}}{\displaystyle f^{2}}=\lim\limits_{f\rightarrow0}
\frac{\displaystyle(8{\rm i}\epsilon\sqrt{d_{1}^{2}+d_{2}^{2}}f^{2}+T_{1}[1])\Phi_{1}}{\displaystyle f^{2}}=
8{\rm i}\epsilon\sqrt{d_{1}^{2}+d_{2}^{2}}\Phi_{1}^{[0]}+T_{1}[1]\Phi_{1}^{[1]}\equiv\Phi_{1}[1],
\end{equation}
where $\Phi_{1}^{[1]}=\displaystyle\frac{1}{2}\frac{\partial^{2}}{\partial f^{2}}|_{f=0}\Phi_{1}$,  we get
a special solution of the
Lax pair (3) and (4) with $u[1]$, $v[1]$ and $\zeta=\zeta_{1}=8{\rm i}\epsilon\sqrt{d_{1}^{2}+d_{2}^{2}}$.
By making use of (\ref{18}) and (\ref{19}), the explicit second-order localized wave solution
can be obtained. Here, we only give the explicit expressions of $u[2]$ and $v[2]$ in the simplest case of
$\alpha=0$. For the case of $\alpha\neq0$, we omit presenting since
the expressions are rather cumbersome to write them down here, although it is not difficult
to verify the validity of the solution by putting them back into Eqs. (1) and (2) using Maple.

(i) $\alpha=0$. By taking $d_{1}=1,d_{2}=3/2$, we calculate that
\begin{equation}\label{32}
u[2]_{rw}=\exp({\displaystyle\frac{13}{4}{\rm i}t})(1+\frac{F_{2}+{\rm i}H_{2}}{D_{2}}),\
v[2]_{rw}=\frac{3}{2}\exp({\displaystyle\frac{13}{4}{\rm i}t})(1+\frac{F_{2}+{\rm i}H_{2}}{D_{2}}),
\end{equation}
where
$$\begin{array}{l}
F_{2}=
-129792\epsilon^{2}x^{4}+10123776\epsilon^{3}tx^{3}-(296120448\epsilon^{4}t^{2}+2530944\epsilon^{2}t^{2}+59904\epsilon^{2})x^{2}
+(3849565824\epsilon^{5}t^{3}
\\~~~~~+98706816\epsilon^{3}t^{3}
+5451264\epsilon^{3}t+7488\epsilon m_{1})x
-24\epsilon(781943058\epsilon^{5}+285610\epsilon+40099644\epsilon^{3})t^{4}\\~~~~~-24\epsilon(3480048\epsilon^{3}+24336\epsilon)t^{2}
+24\epsilon(156\sqrt{13}n_{1}
-6084\epsilon m_{1})t+2304\epsilon^{2},\\
G_{2}=
-843648t\epsilon^{2}x^{4}+65804544\epsilon^{3}t^{2}x^{3}-(1924782912\epsilon^{4}t^{3}-389376\epsilon^{2}t+3744n_{1}\epsilon \sqrt{13}+5483712\epsilon^{2}t^{3})x^{2}
\\~~~~~+(25022177856\epsilon^{5}t^{4}+213864768\epsilon^{3}t^{4}+5061888\epsilon^{3}t^{2}+48672\epsilon m_{1}t+146016\sqrt{13}\epsilon^{2}n_{1}t)x
-24\epsilon(371293\epsilon\\~~~~~+5082629877\epsilon^{5}+86882562\epsilon^{3})t^{5}-24\epsilon(17576\epsilon
+10281960\epsilon^{3})t^{3}+24\epsilon(507\sqrt{13}n_{1}
-59319\sqrt{13}\epsilon^{2}n_{1}\\~~~~~-39546\epsilon m_{1})t^{2}+74880\epsilon^{2}t-288\sqrt{13}\epsilon n_{1},\\
D_{2}=140608\epsilon^{2}x^{6}-16451136\epsilon^{3}tx^{5}+((801992880\epsilon^{4}+1370928\epsilon^{2})t^{2}
+32448\epsilon^{2})x^{4}
-((106932384\epsilon^{3}\\~~~~~+20851814880\epsilon^{5})t^{3}-843648\epsilon^{3}t-8112\epsilon m_{1})x^{3}
+((304957792620\epsilon^{6}+4455516\epsilon^{2}+3127772232\epsilon^{4})t^{4}\\~~~~~-(123383520\epsilon^{4}
+632736\epsilon^{2})t^{2}+(12168\sqrt{13}\epsilon n_{1}
-474552\epsilon^{2}m_{1})t+22464\epsilon^{2})x^{2}-((173765124\epsilon^{3}\\~~~~~+40661039016\epsilon^{5}
+2378670782436\epsilon^{7})t^{5}-(2887174368\epsilon^{5}
-8225568\epsilon^{3})t^{3}-(9253764\epsilon^{3}m_{1}-79092\epsilon m_{1}\\~~~~~-474552\sqrt{13}\epsilon^{2}n_{1})t^{2}+1654848\epsilon^{3}t+1872\epsilon m_{1})x
+(4826809\epsilon^{2}+7730680042917\epsilon^{8}+198222565203\epsilon^{6}\\~~~~~+1694209959\epsilon^{4})t^{6}
+(400996440\epsilon^{4}+3084588\epsilon^{2}
-20330519508\epsilon^{6})t^{4}+(4626882\sqrt{13}\epsilon^{3}n_{1}\\~~~~~-13182\sqrt{13}\epsilon n_{1} +1542294\epsilon^{2}m_{1}-60149466\epsilon^{4}m_{1})t^{3}+(43975152\epsilon^{4}
+267696\epsilon^{2})t^{2}
+(133848\epsilon^{2}m_{1}\\~~~~~-2808\sqrt{13}\epsilon n_{1} )t+117m_{1}^{2}+117n_{1}^{2}+576\epsilon^{2}.
\end{array}
$$
Then $u[2]_{rw}$ is merely proportional to $v[2]_{rw}$ with the radio of $2/3$, the maximum value of
them with $m_{1}=0$ and $n_{1}=0$ are five times more than their each average crest.
Thus, the above solution is the vector generalization of the second-order rogue wave solution to the
decoupled Hirota equation.

(ii) $\alpha\neq0$, $d_{1}\neq0$ and  $d_{2}=0$. Hence we arrive at the interactional solution between two dark-bright
solitons and a second-order rogue wave. Fig. 7(a) displays two dark solitons together with a
fundamental second-order rogue wave, Fig. 7(b) shows  two bright solitons coexist with a fundamental
second-order rogue wave. The explicit collision processes
are exhibited in Figs. 8 and 9. It is shown that the interactional process is also elastic,
the amplitudes and velocities of the two dark and bright solitons remain unchanged after the collision.
When by decreasing the value of $\alpha$, the two dark-bright solitons and the fundamental second-order
rogue wave separate, see Fig. 10. At this moment, when the two bright solitons and the second-order rogue wave divide,
the second-order rogue wave in $v$ component are also unobservable for the same reason as the first-order case.
When setting $s_{1}\neq0$, the fundamental second-order rogue wave can split into three first-order rogue waves, see Fig. 11.

(iii) $\alpha\neq0$, $d_{1}\neq0$ and  $d_{2}\neq0$. Here, the interactional solutions between two Akhmediev
breathers and a second-order rogue wave can be presented, see Figs. 12-14. We see that by increasing the
value of $\alpha$, the two Akhmediev breathers and the second-order rogue wave merge, see Fig. 12.
While by decreasing the value of $\alpha$, the two parallel Akhmediev breathers and the second-order rogue separate, see
Fig. 13. Meanwhile, when taking $s_{1}\neq0$, the fundamental second-order rogue wave
can split into three first-order rogue waves, see Fig. 14.

\section{Conclusion}
In summary, we present some novel localized wave solutions of the coupled Hirota equations with
high-order nonlinear effects like the third dispersion, self-steepening and inelastic Raman scattering terms.
By using the Taylor series expansion coefficients of a special solution (\ref{27}) to the Lax pair equations
(\ref{03}) and (\ref{04}) with a periodic seed solution and a fixed spectral parameter,
a generalized Darboux transformation of the coupled Hirota equations is constructed. As application,
the interesting interactions between rogue waves and the other nonlinear waves such as dark-bright solitons
and Akhmediev breathers in the coupled Hirota equations are illustrated through some figures.
Several free parameters such as $\alpha$, $d_{i}~(i=1,2)$ and $s_{1}$ play an important part to affect the dynamic structures of the
localized waves. (1) When $\alpha=0$, the first- and  the second-order localized wave solutions
are the vector generalization of the corresponding rogue wave solutions to the decoupled Hirota equation.
(2) When $\alpha\neq0$, $d_{1}\neq0$ and $d_{2}=0$, the first- and the second-order dark-bright-rogue wave solutions
are reached. (3) When $\alpha\neq0$, $d_{1}\neq0$ and $d_{2}\neq0$, the first- and the second-order breather-rogue
wave solutions are derived. Moreover, by increasing the value of $\alpha$,
the rogue waves and the dark-bright solitons or breathers merge, by decreasing the value of $\alpha$
they separate, and by taking $s_{1}\neq0$, the fundamental second-order rogue wave can split into three
first-order rogue waves. 
Our results can be seen as the generalization of the work done by
Baronio et al. \cite{27} to the complex coupled  system with high-order nonlinear terms.

Besides, on the one hand, continuing the generalized DT process, the more complicated localized wave solutions can be generated,
which may possess more abundant striking dynamics. On the other hand, the interactions between 
rogue waves and cnoidal waves have recently been reported in the NLS equation \cite{co}, there are many possibilities 
to be observed in the coupled Hirota equations. 
Both of them are interesting and we will investigate in the future paper.
The results in the present paper further reveal the intriguing dynamic distributions of localized waves in the
coupled Hirota equations with high-order nonlinear terms,
and we hope our results can be verified in real experiments in the future.
\begin{figure}[!h]
\centering
\renewcommand{\figurename}{{\bf Fig.}}
{\includegraphics[height=4cm,width=6cm]{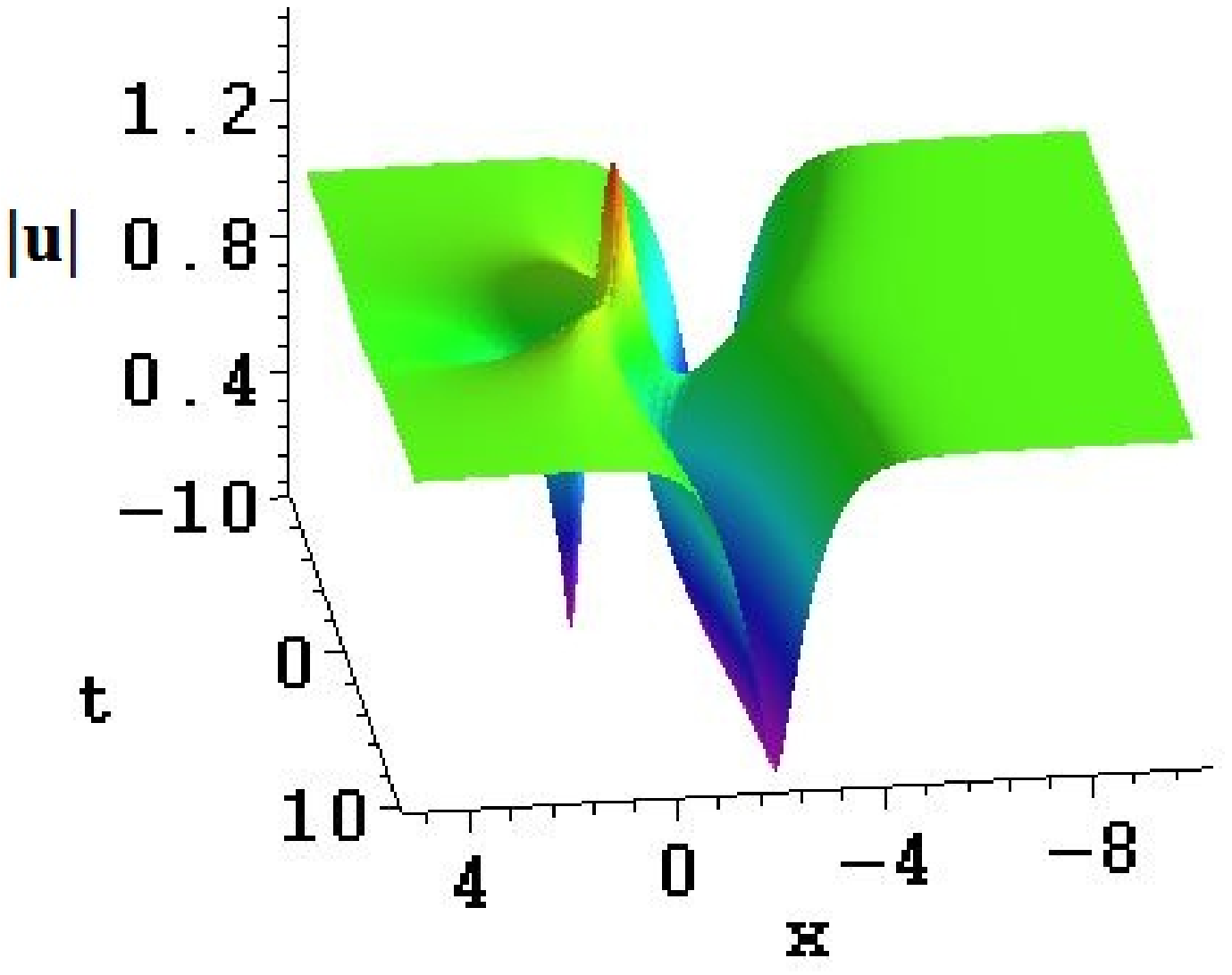}}
{\includegraphics[height=4cm,width=6cm]{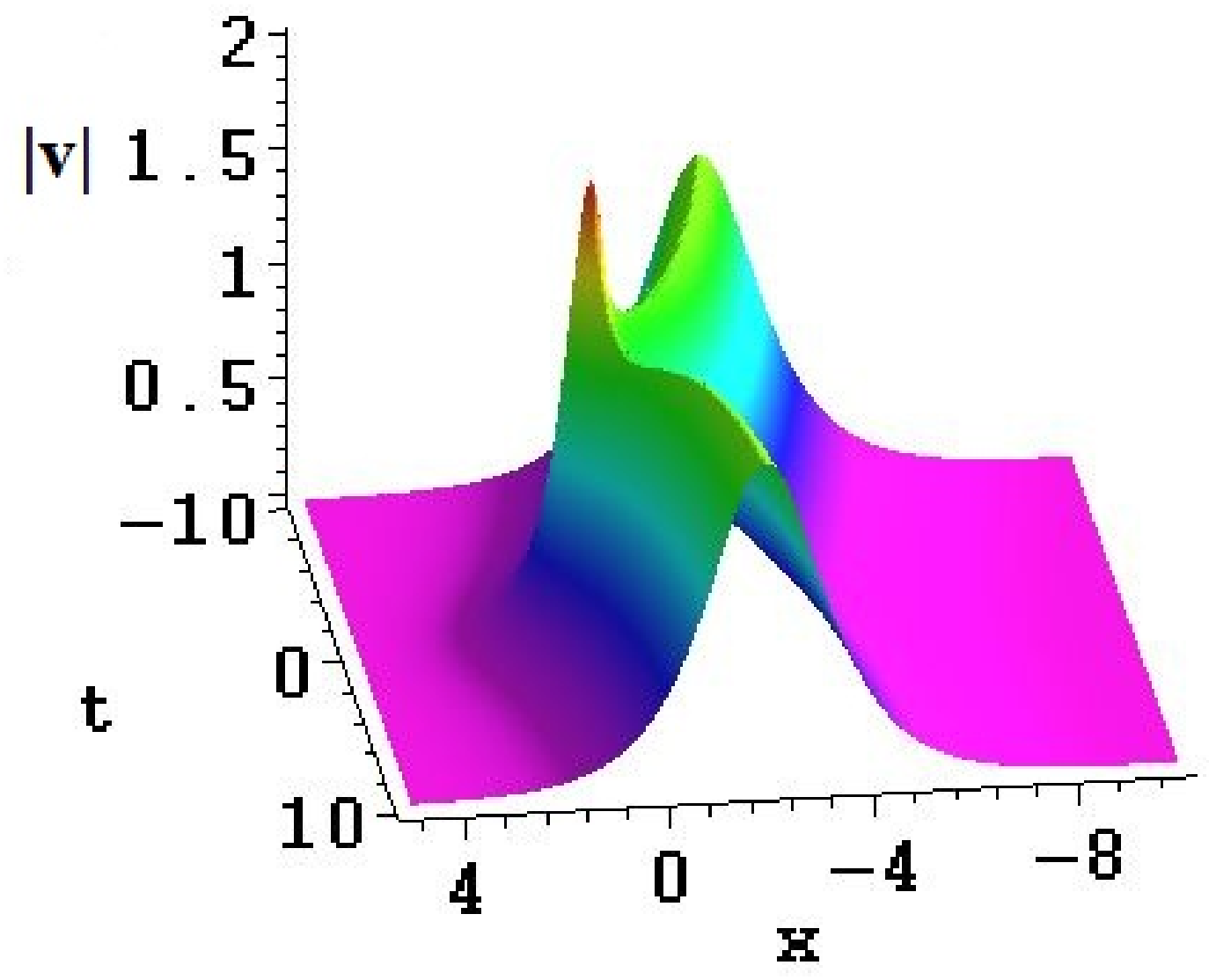}}
\begin{center}
\hskip 1cm $(\rm{a})$ \hskip 5cm $(\rm{b})$
\end{center}
\caption{Evolution plot of the dark-bright-rogue wave in coupled Hirota equations with the parameters chosen by
$\epsilon=1/100,\alpha=10,d_{1}=1,d_{2}=0$. (a) a dark soliton merge with a rogue wave
in $u$ component; (b) a bright soliton merge with a rogue wave in $v$ component.}
\end{figure}

\begin{figure}
\centering
\renewcommand{\figurename}{{\bf Fig.}}
{\includegraphics[height=4cm,width=4cm]{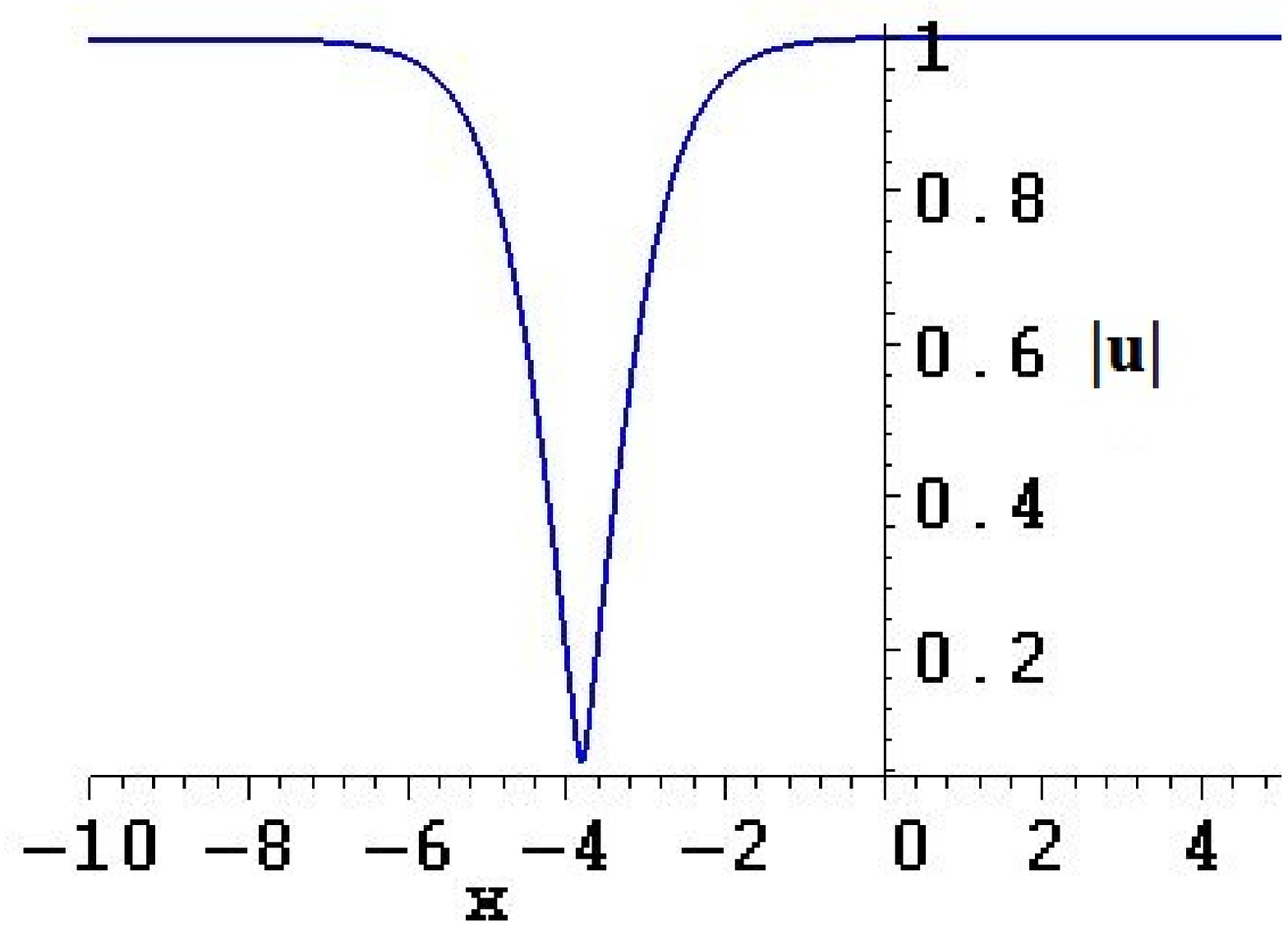}}
{\includegraphics[height=4cm,width=4cm]{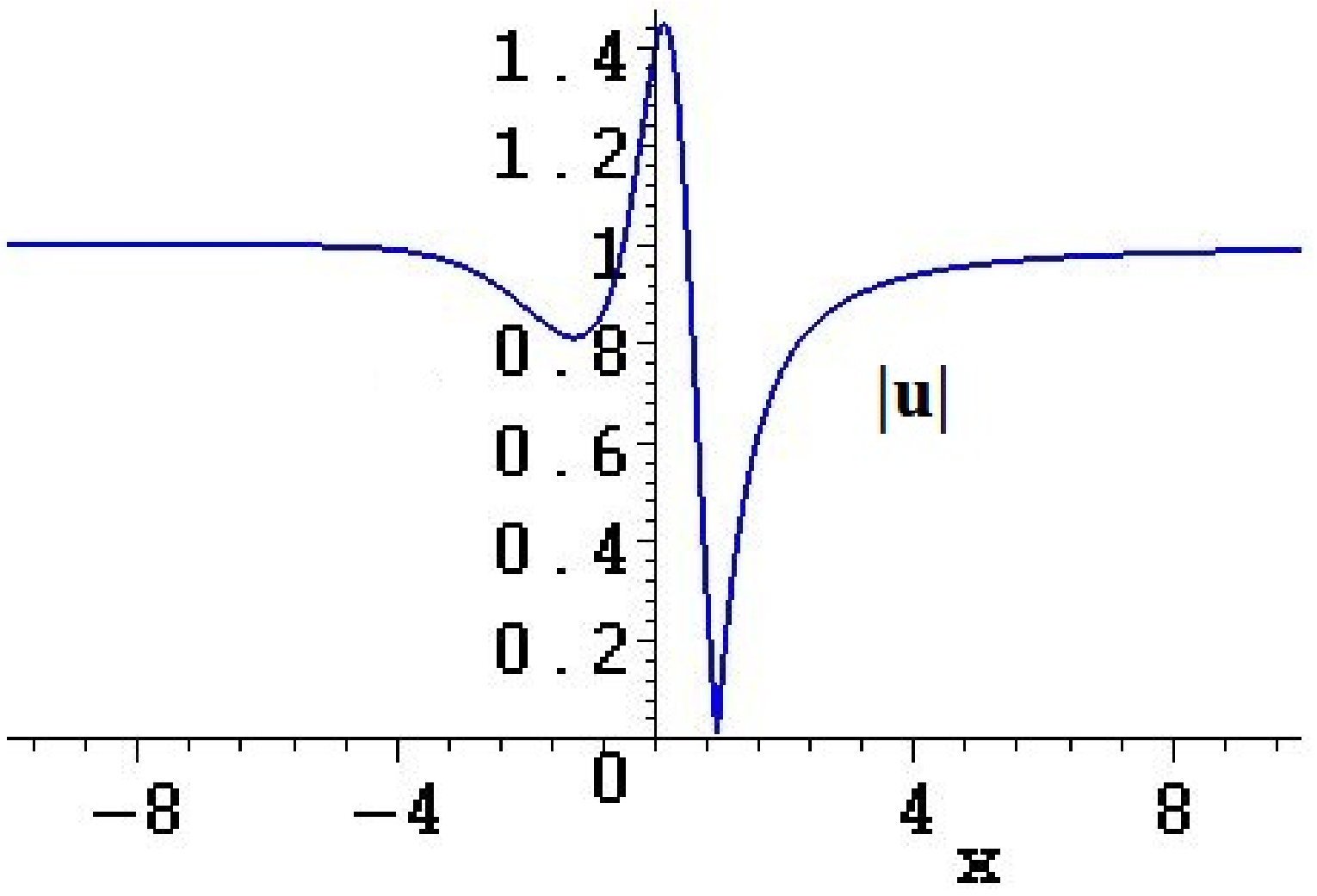}}
{\includegraphics[height=4cm,width=4cm]{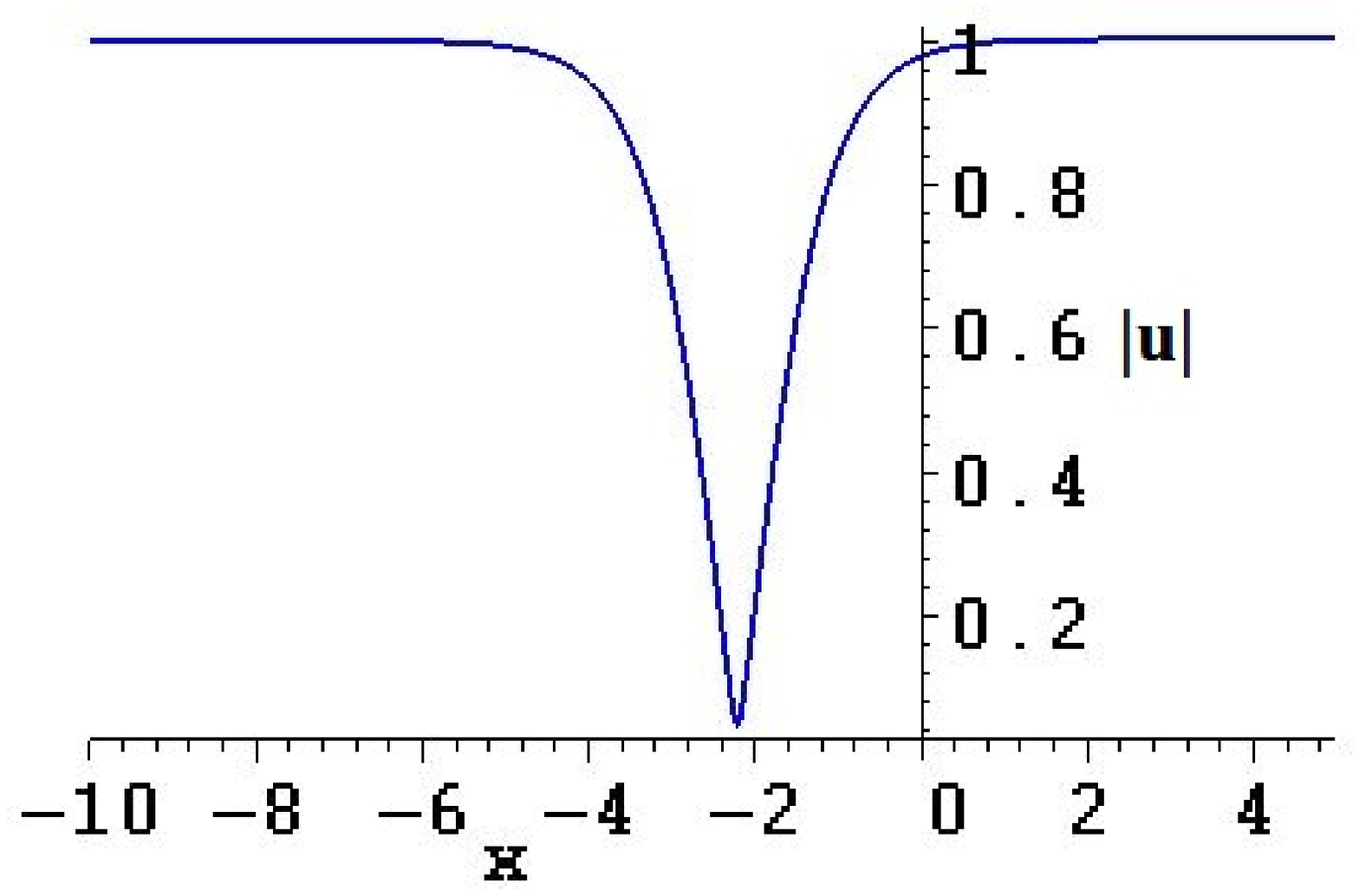}}
\begin{center}
\hskip 1cm $(\rm{a})$ \hskip 4cm $(\rm{b})$ \hskip 4cm $(\rm{c})$
\end{center}
\caption{Plane evolution plot of the interactional process between a right-going dark soliton and a rogue wave
in Fig. 1(a) at: (a) $t=-20$; (b) $t=0$; (c) $t=20$. The collision process is elastic, the amplitude and velocity of the
dark soliton are unchanged after the collision.}
\centering
\renewcommand{\figurename}{{\bf Fig.}}
{\includegraphics[height=4cm,width=4cm]{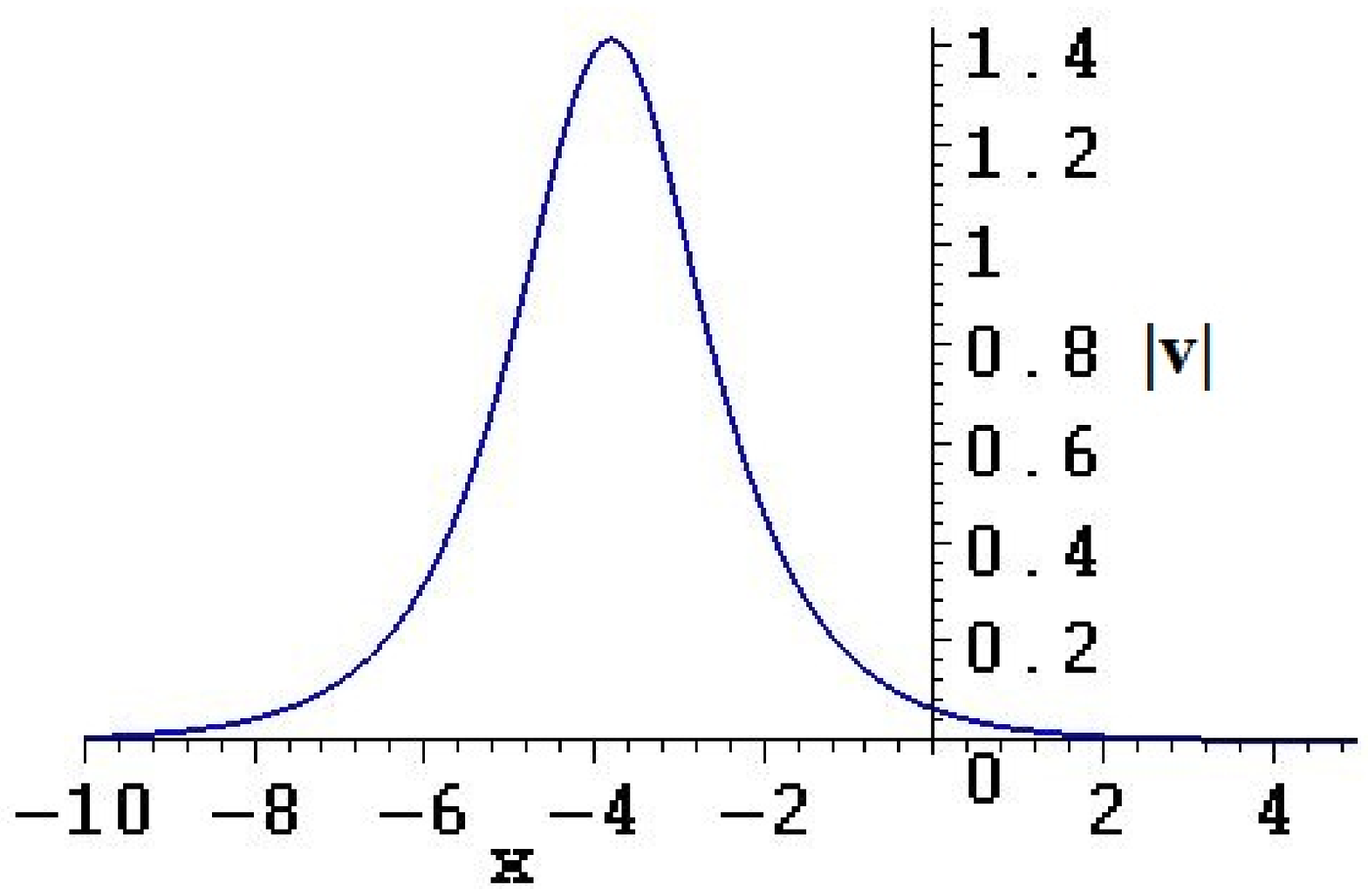}}
{\includegraphics[height=4cm,width=4cm]{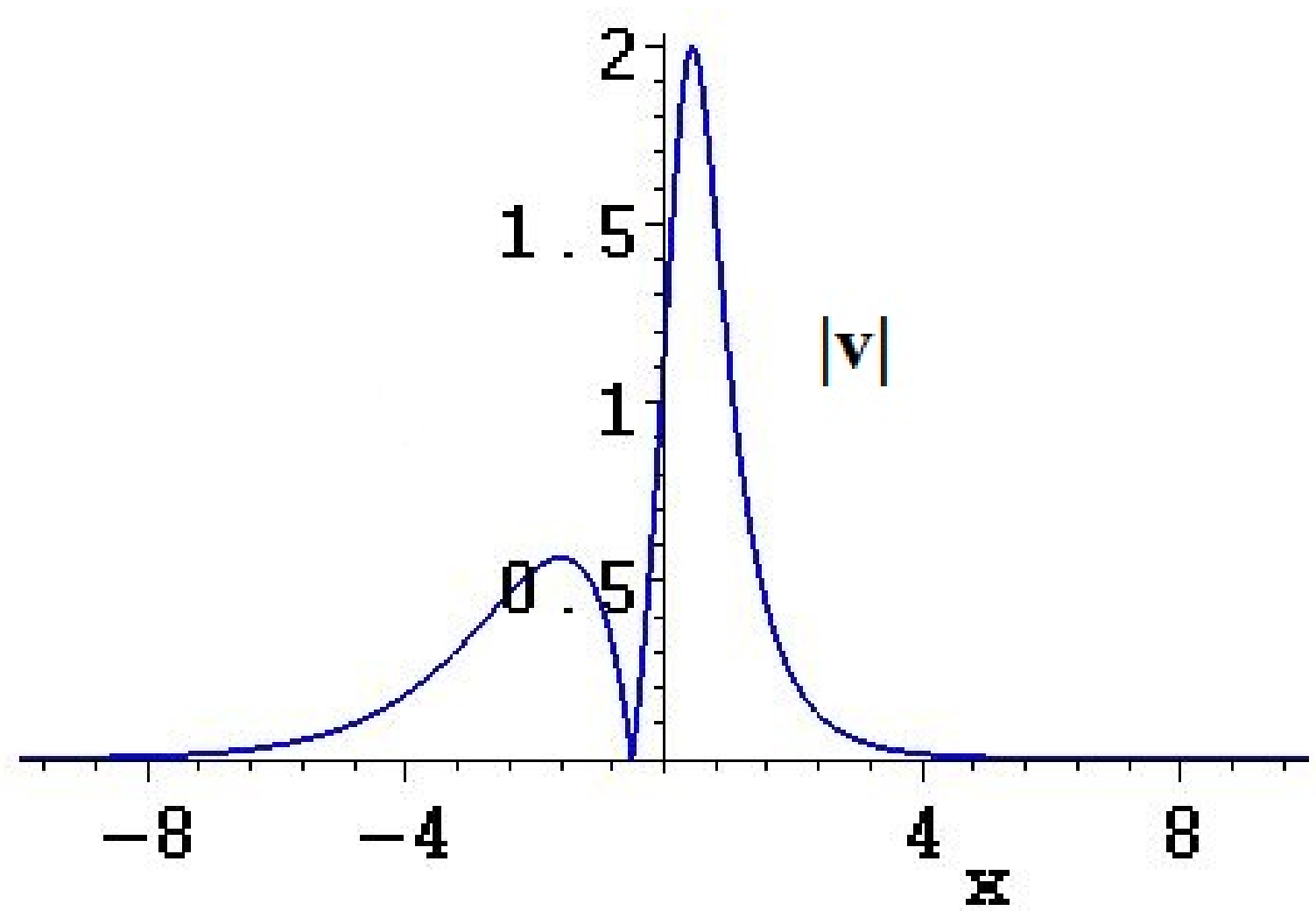}}
{\includegraphics[height=4cm,width=4cm]{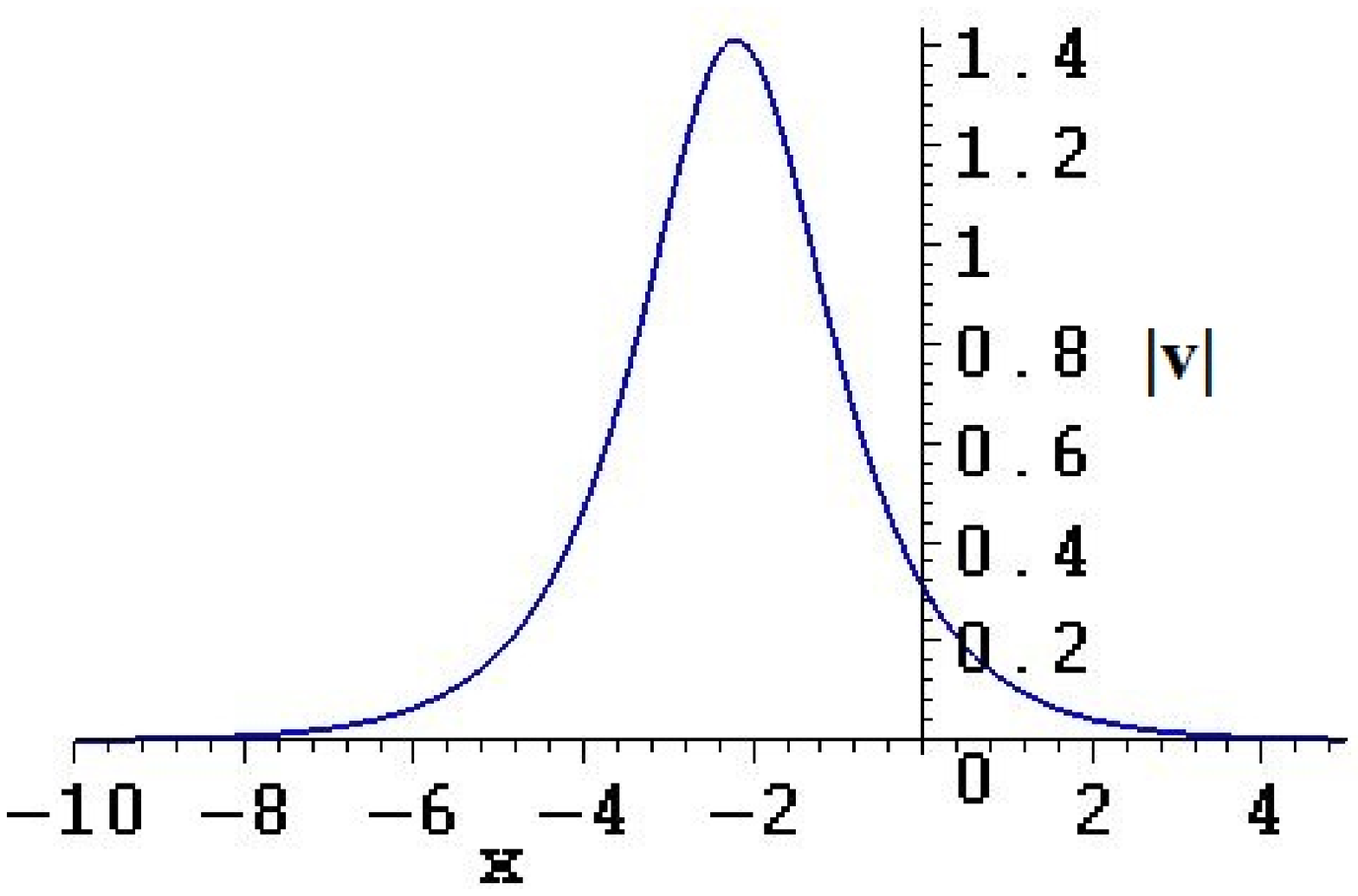}}
\begin{center}
\hskip 1cm $(\rm{a})$ \hskip 4cm $(\rm{b})$ \hskip 4cm $(\rm{c})$
\end{center}
\caption{Plane evolution plot of the interactional process between a right-going bright soliton and a rogue wave
in Fig. 1(b) at: (a) $t=-20$; (b) $t=0$; (c) $t=20$. The collision process is elastic, the amplitude and velocity of the
bright soliton are unchanged after the collision.}
\centering
\renewcommand{\figurename}{{\bf Fig.}}
{\includegraphics[height=4cm,width=6cm]{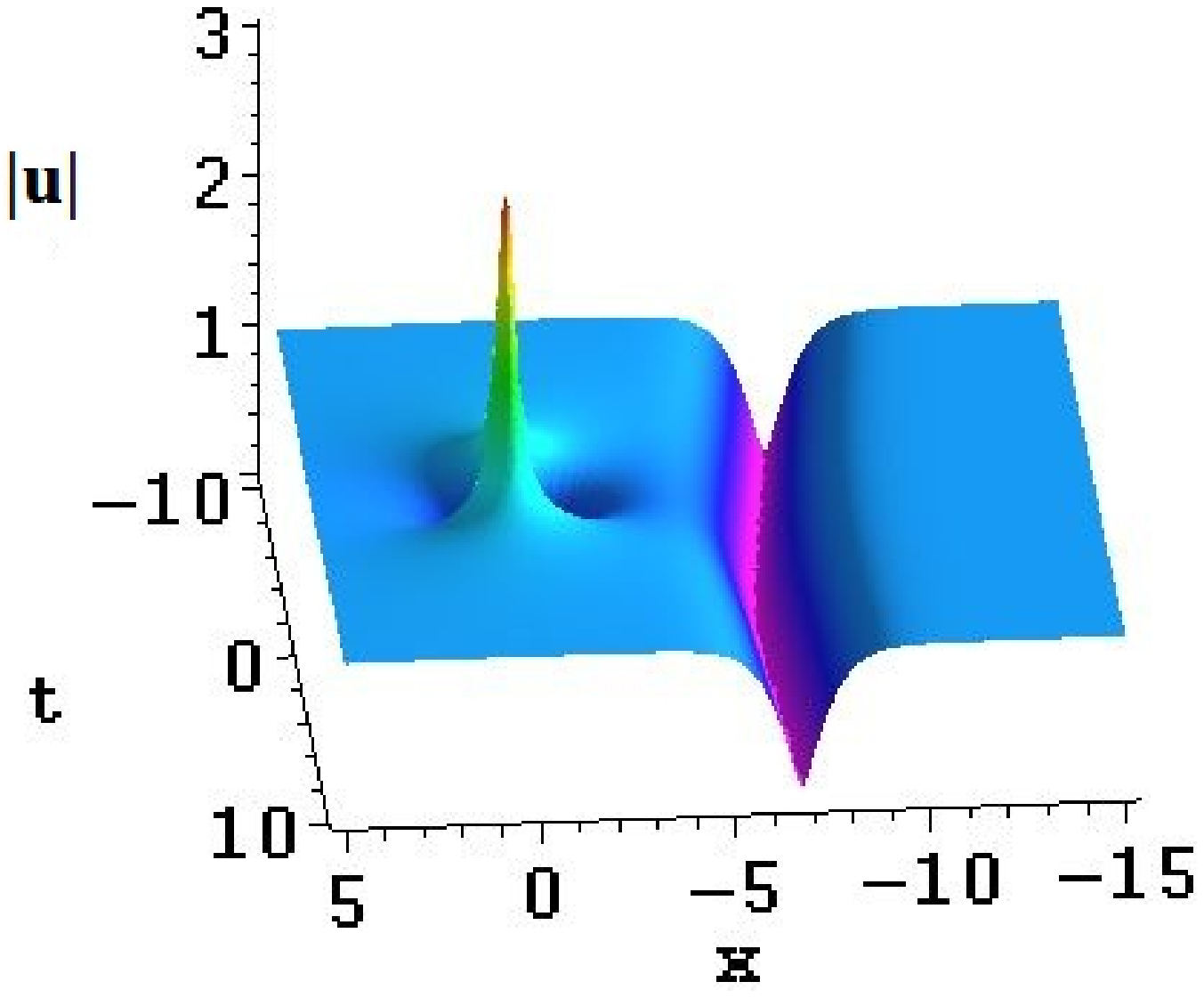}}
{\includegraphics[height=4cm,width=6cm]{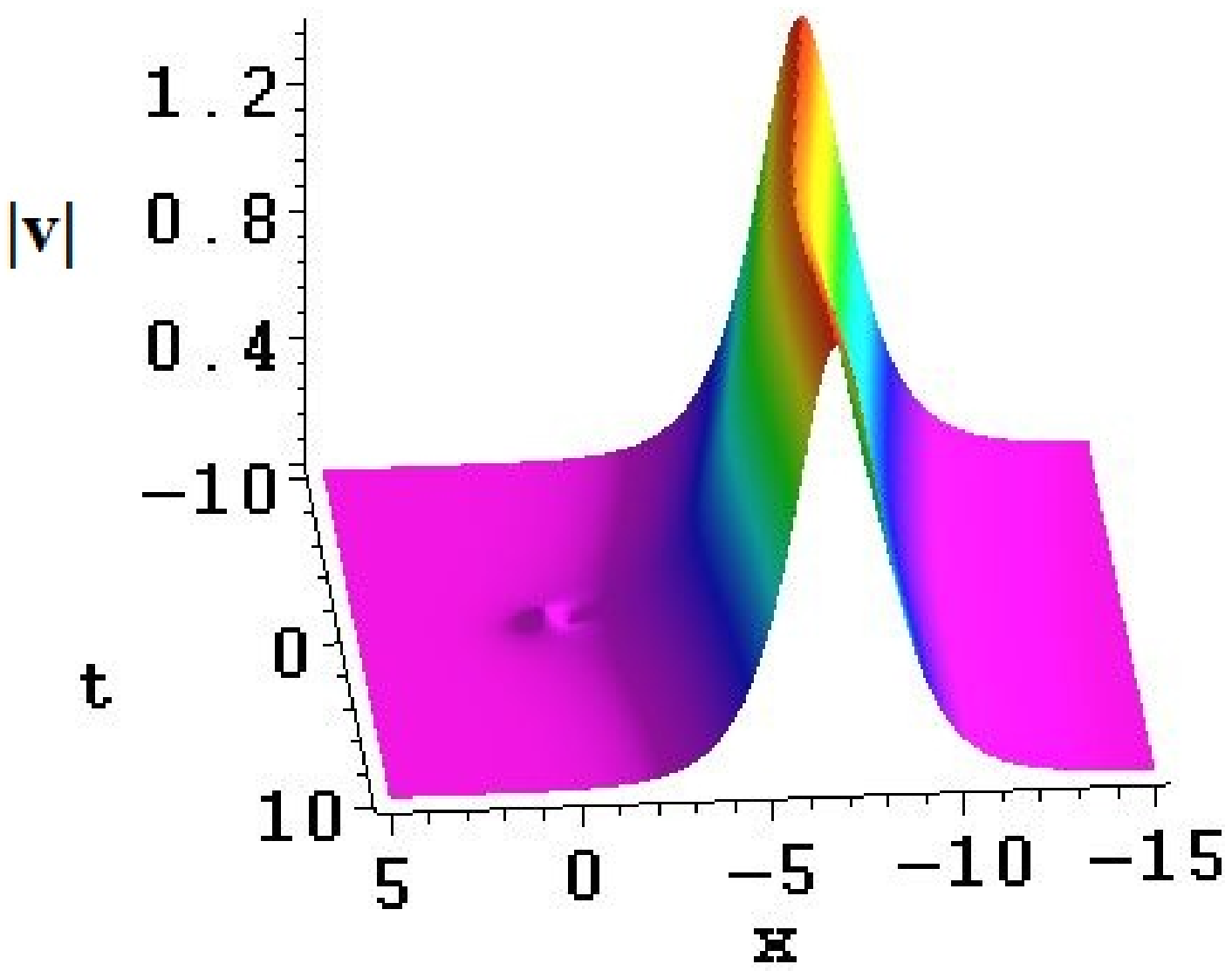}}
\begin{center}
\hskip 1cm $(\rm{a})$ \hskip 5cm $(\rm{b})$
\end{center}
\caption{Evolution plot of the dark-bright-rogue wave in coupled Hirota equations with the parameters chosen by
$\epsilon=1/100,\alpha=1/10,d_{1}=1,d_{2}=0$.
(a) the dark soliton and the rogue wave separate in $u$ component;
(b) the bright soliton and the rogue wave separate in $v$ component.
The rogue wave in $v$ component is difficult to be seen for its zero-amplitude background wave.}
\centering
\renewcommand{\figurename}{{\bf Fig.}}
{\includegraphics[height=4cm,width=6cm]{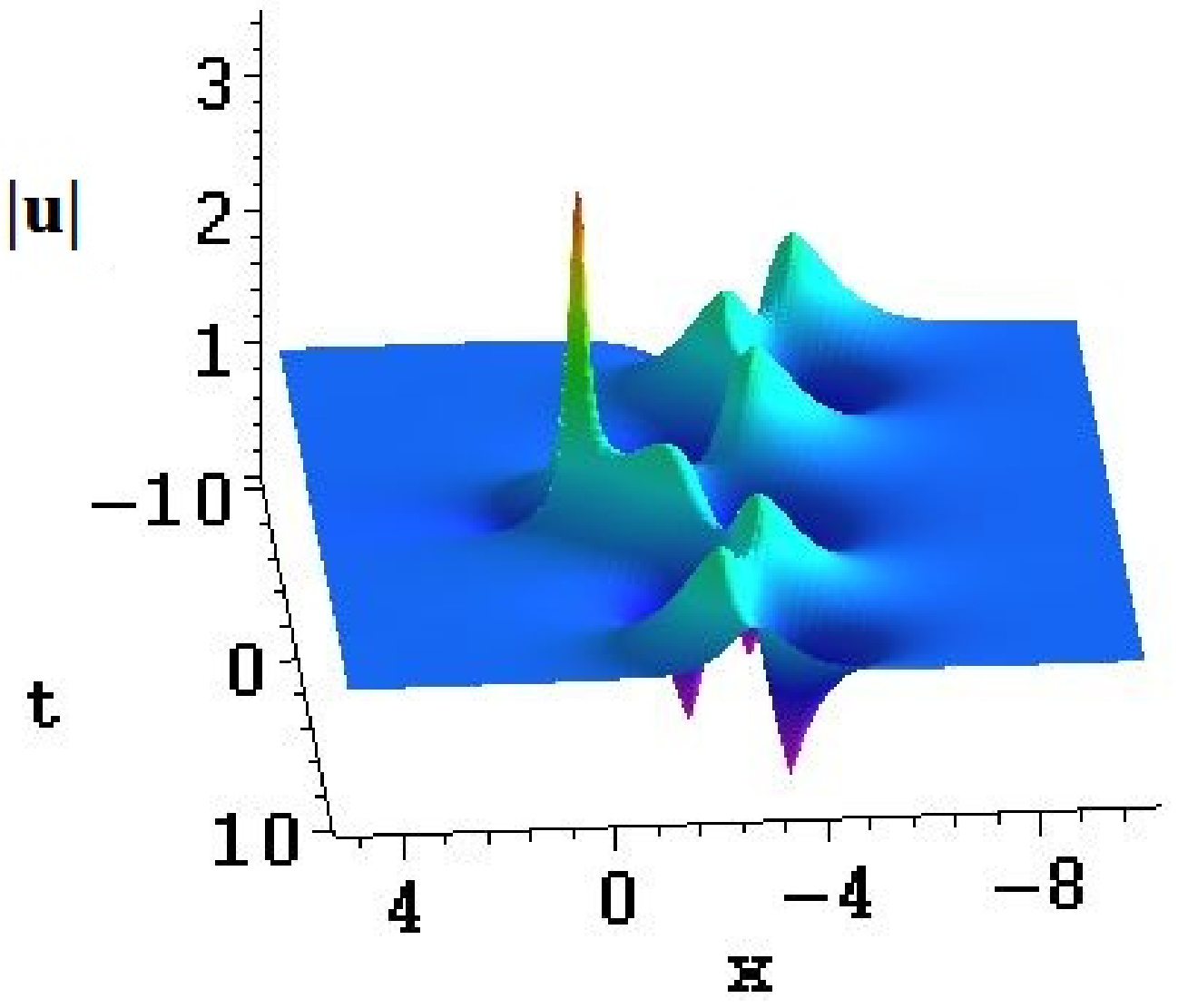}}
{\includegraphics[height=4cm,width=6cm]{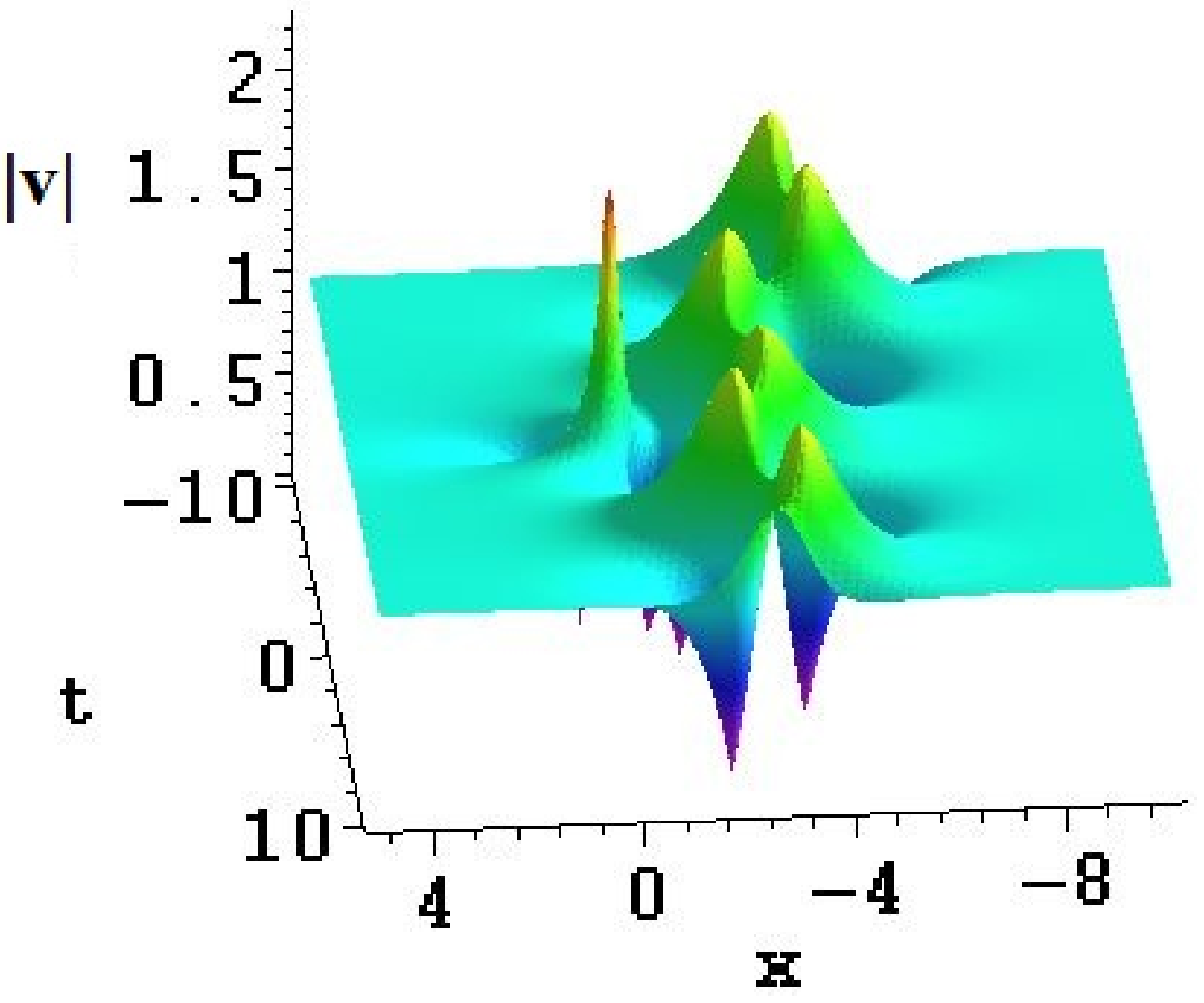}}
\begin{center}
\hskip 1cm $(\rm{a})$ \hskip 5cm $(\rm{b})$
\end{center}
\caption{Evolution plot of the breather-rogue wave in coupled Hirota equations with the parameters chosen by
$\epsilon=1/100,\alpha=1,d_{1}=1,d_{2}=1$.
(a) an Akhmediev breather merge with a rogue wave in $u$ component; (b) an Akhmediev breather merge with
a rogue wave in $v$ component.}
\end{figure}

\begin{figure}
\centering
\renewcommand{\figurename}{{\bf Fig.}}
{\includegraphics[height=4cm,width=6cm]{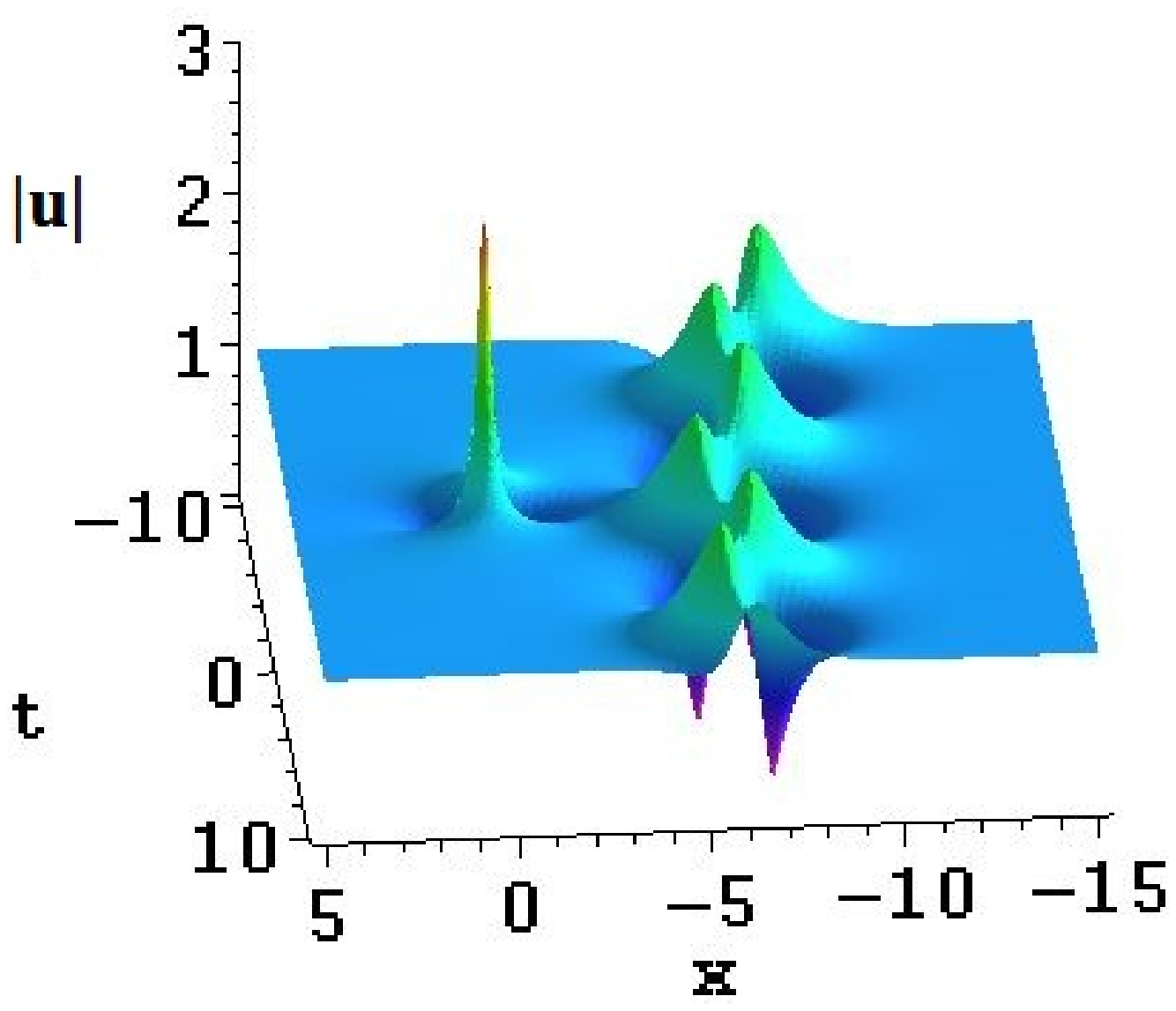}}
{\includegraphics[height=4cm,width=6cm]{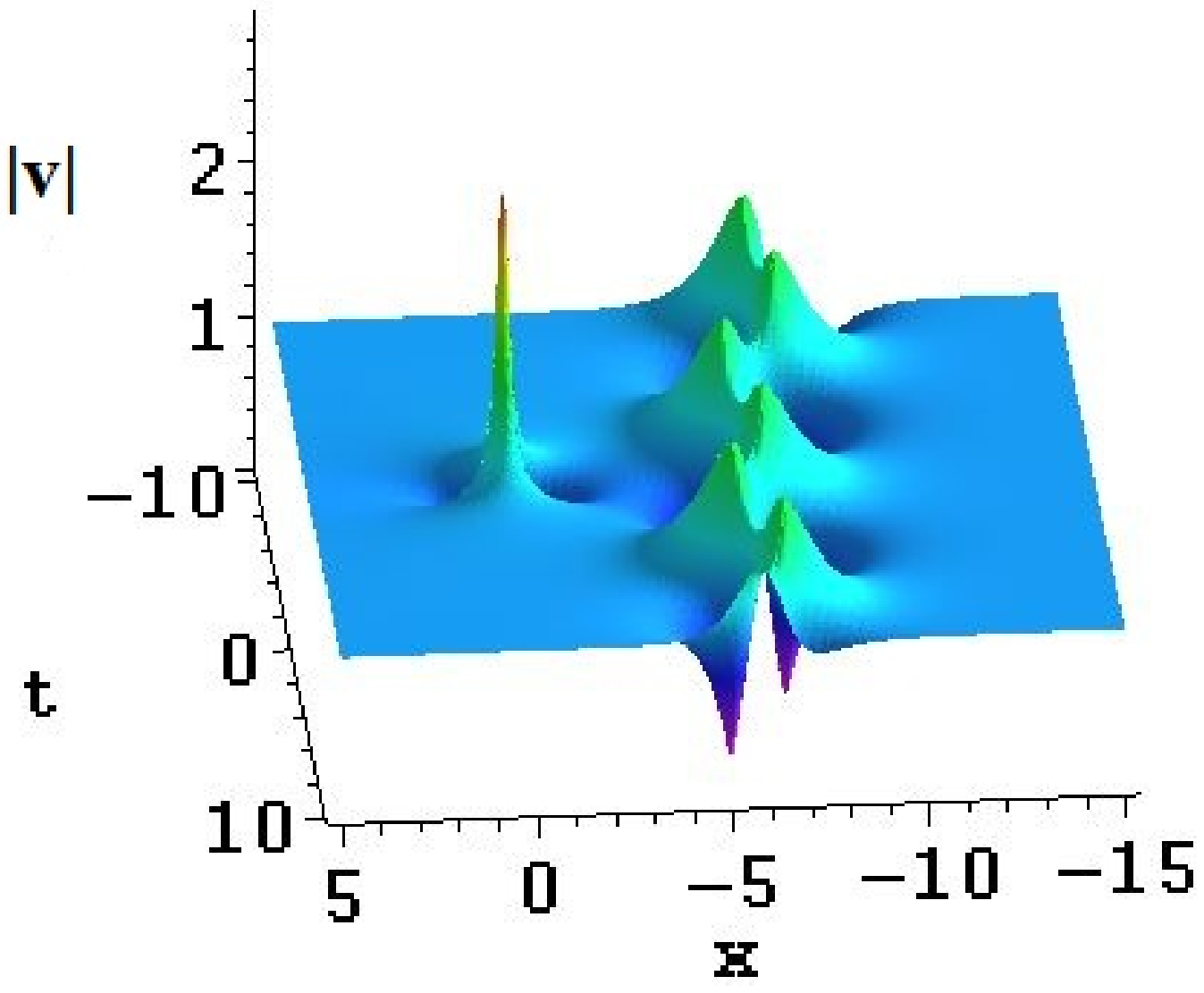}}
\begin{center}
\hskip 1cm $(\rm{a})$ \hskip 5cm $(\rm{b})$
\end{center}
\caption{Evolution plot of the breather-rogue wave in coupled Hirota equations with the parameters chosen by
$\epsilon=1/100,\alpha=1/100,d_{1}=1,d_{2}=1$.
(a) the Akhmediev breather and the rogue wave separate in $u$ component; (b)
the Akhmediev breather and the rogue wave separate in $v$ component.}
\centering
\renewcommand{\figurename}{{\bf Fig.}}
{\includegraphics[height=4cm,width=6cm]{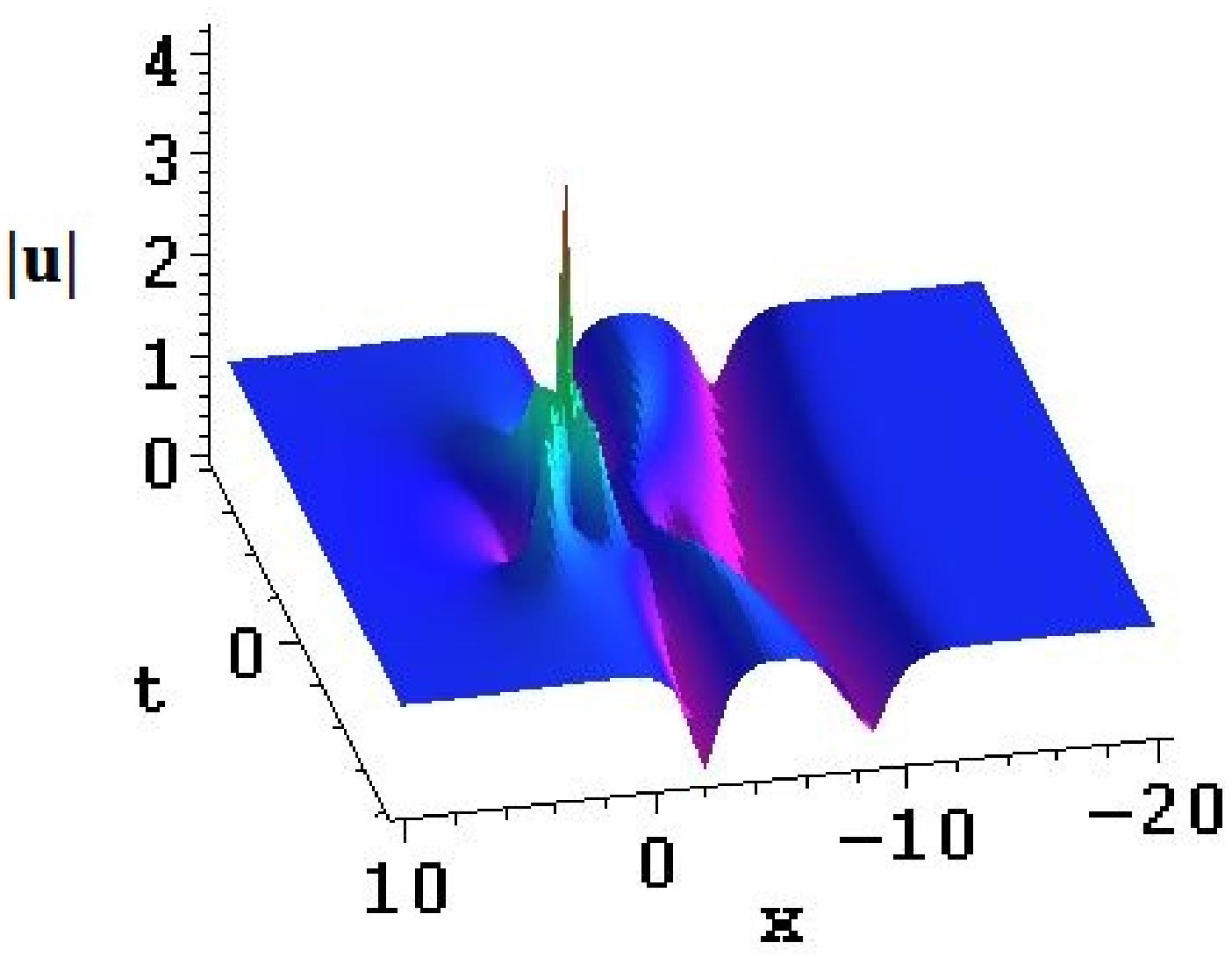}}
{\includegraphics[height=4cm,width=6cm]{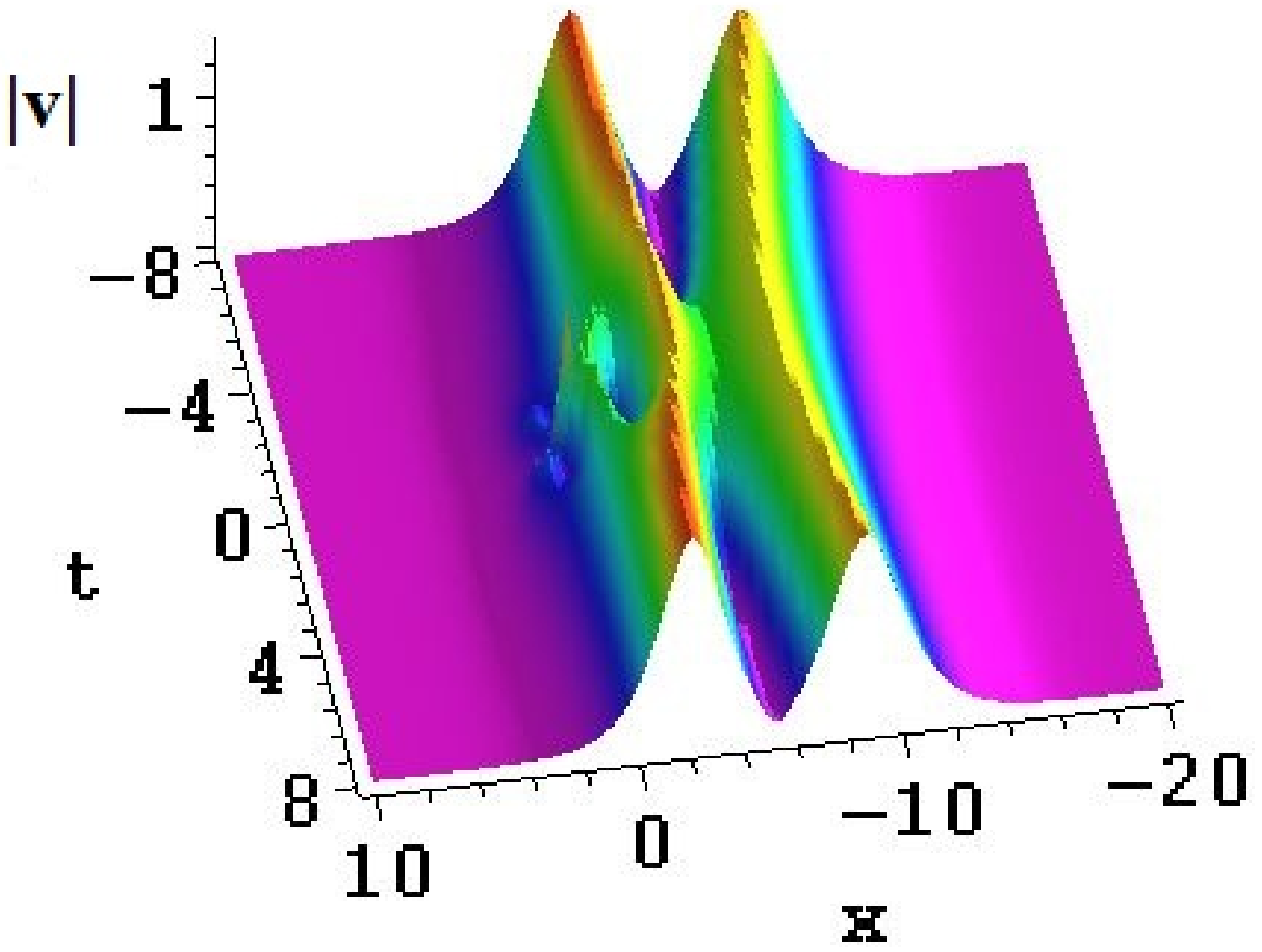}}
\begin{center}
\hskip 1cm $(\rm{a})$ \hskip 5cm $(\rm{b})$
\end{center}
\caption{Evolution plot of the second-order dark-bright-rogue wave in coupled Hirota equations with the parameters chosen by
$\epsilon=1/100,\alpha=1,d_{1}=1,d_{2}=0,m_{1}=0,n_{1}=0$. (a) two dark solitons merge with a fundamental second-order rogue wave
in $u$ component; (b) two bright solitons merge with a fundamental second-order rogue wave
in $v$ component. }
\centering
\renewcommand{\figurename}{{\bf Fig.}}
{\includegraphics[height=4cm,width=4cm]{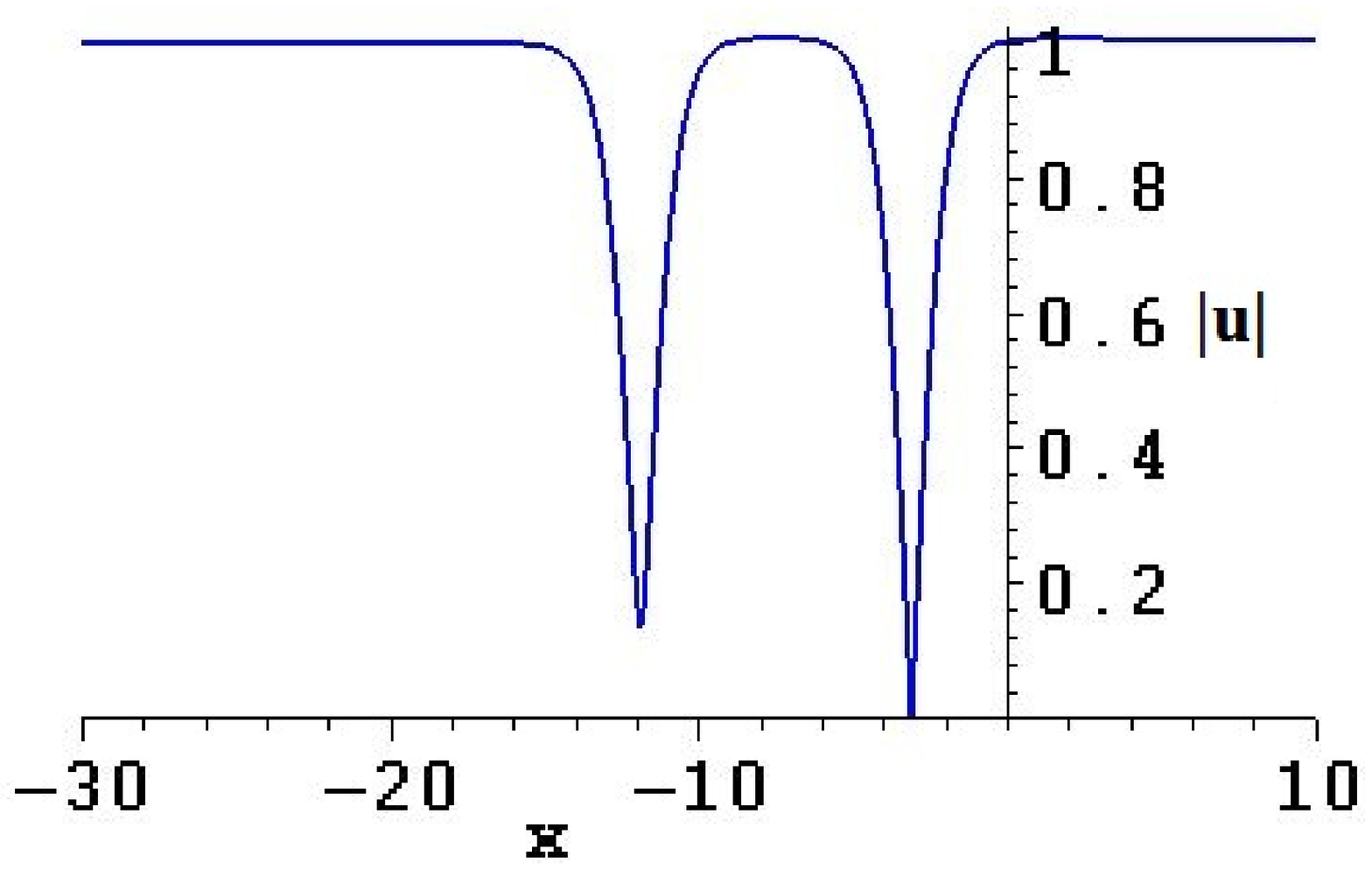}}
{\includegraphics[height=4cm,width=4cm]{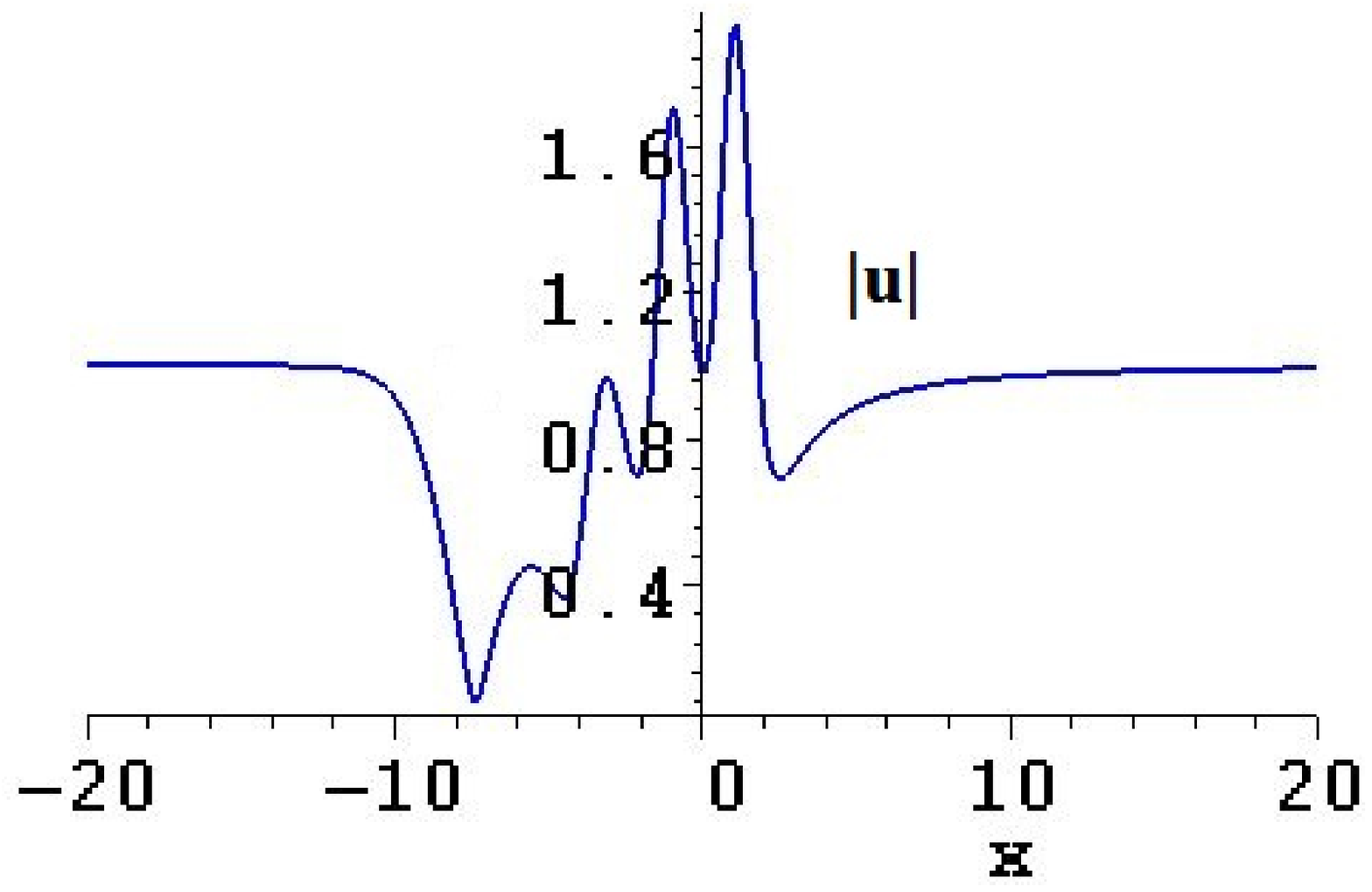}}
{\includegraphics[height=4cm,width=4cm]{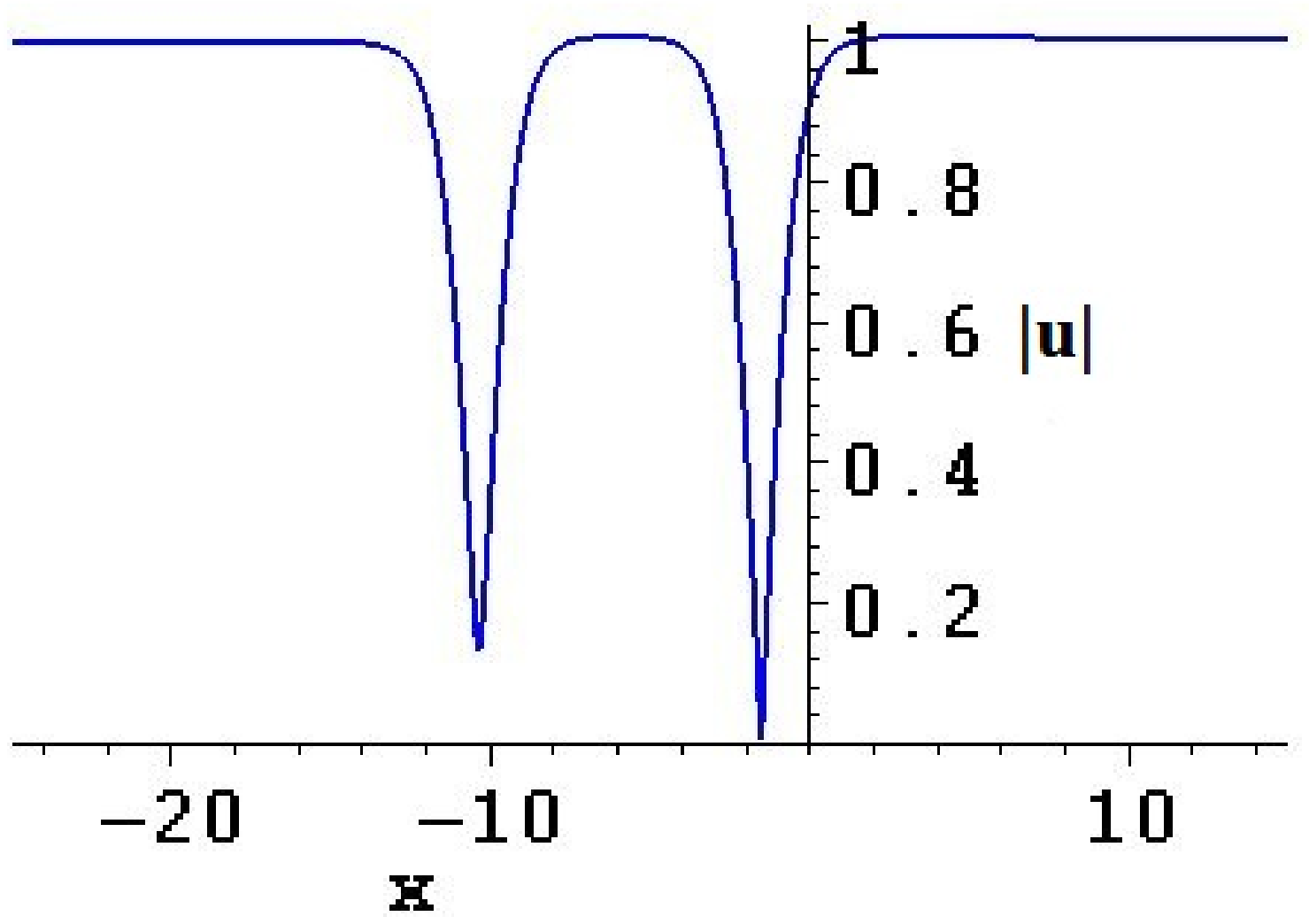}}
\begin{center}
\hskip 1cm $(\rm{a})$ \hskip 4cm $(\rm{b})$ \hskip 4cm $(\rm{c})$
\end{center}
\caption{Plane evolution plot of the interactional process between two dark solitons and
a fundamental second-order rogue wave
in Figs. 7(a) at: (a) $t=-20$; (b) $t=1$; (c) $t=20$. The collision process is elastic, the amplitude and velocity of the
dark solitons are unchanged after the collision.}
\end{figure}

\begin{figure}
\centering
\renewcommand{\figurename}{{\bf Fig.}}
{\includegraphics[height=4cm,width=4cm]{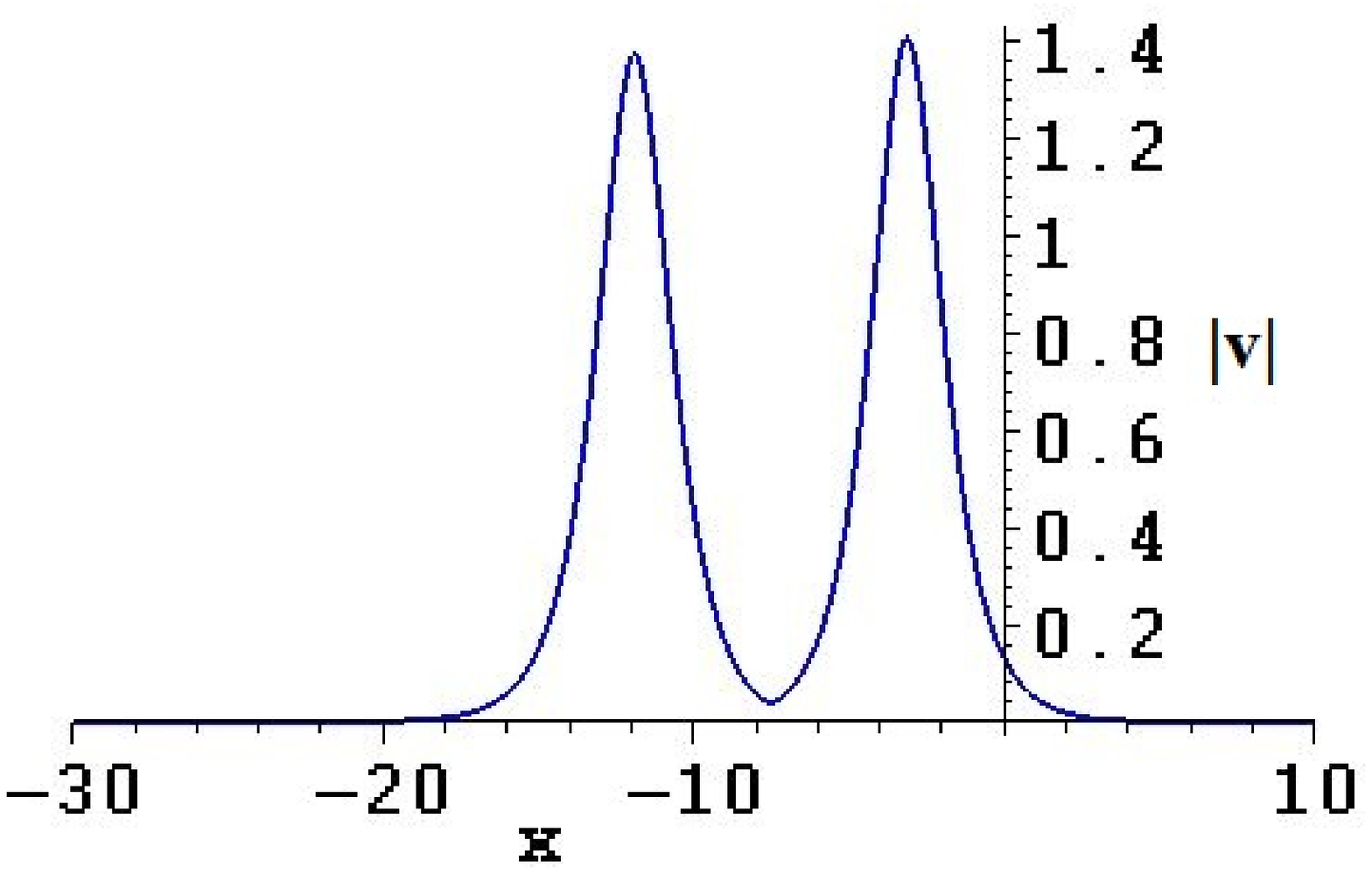}}
{\includegraphics[height=4cm,width=4cm]{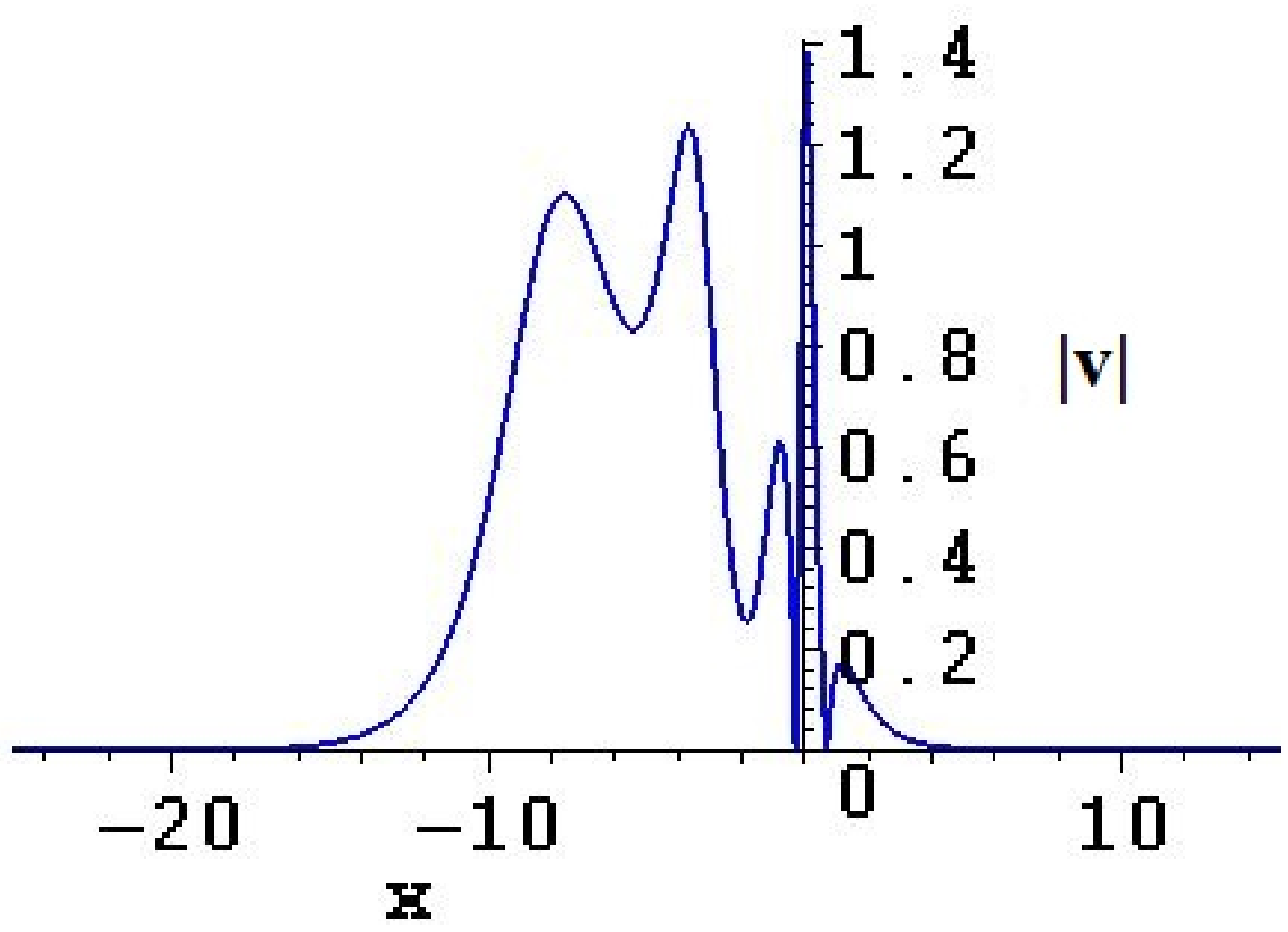}}
{\includegraphics[height=4cm,width=4cm]{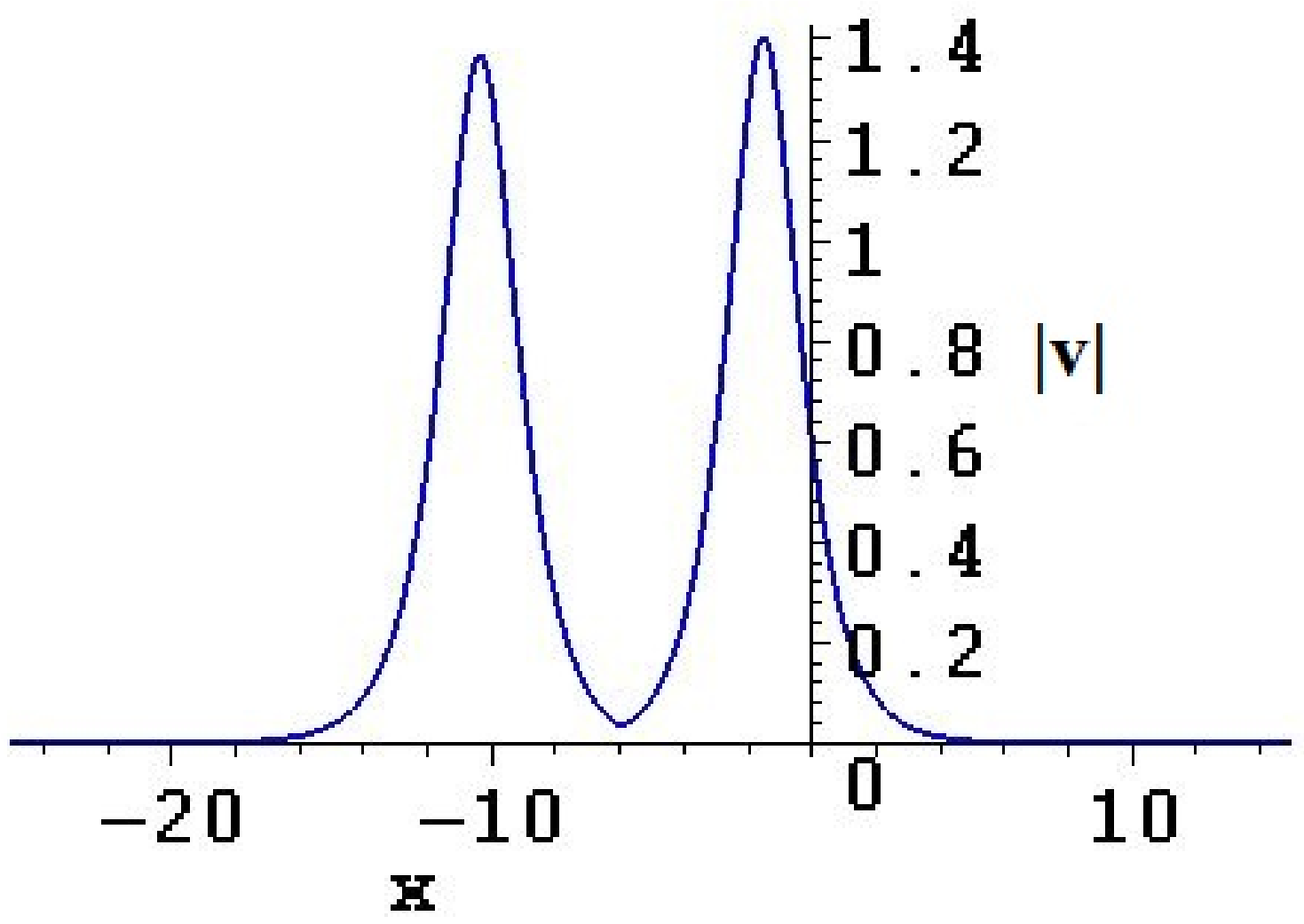}}
\begin{center}
\hskip 1cm $(\rm{a})$ \hskip 4cm $(\rm{b})$ \hskip 4cm $(\rm{c})$
\end{center}
\caption{Plane evolution plot of the interactional process between two bright solitons and
a fundamental second-order rogue wave
in Figs. 7(b) at: (a) $t=-20$; (b) $t=0$; (c) $t=20$. The collision process is elastic, the amplitude and velocity of the
bright solitons are unchanged after the collision.}
\centering
\renewcommand{\figurename}{{\bf Fig.}}
{\includegraphics[height=4cm,width=6cm]{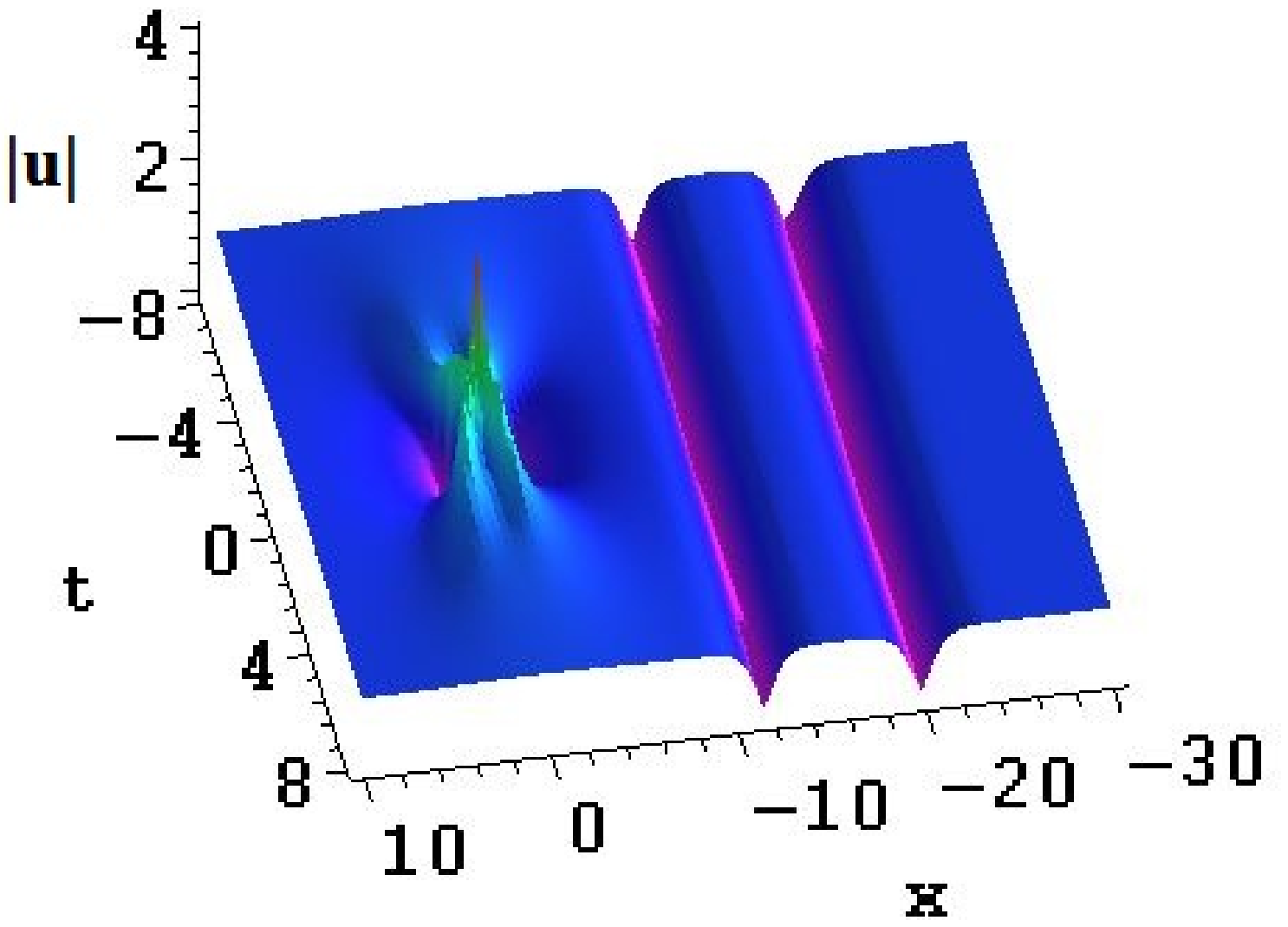}}
{\includegraphics[height=4cm,width=6cm]{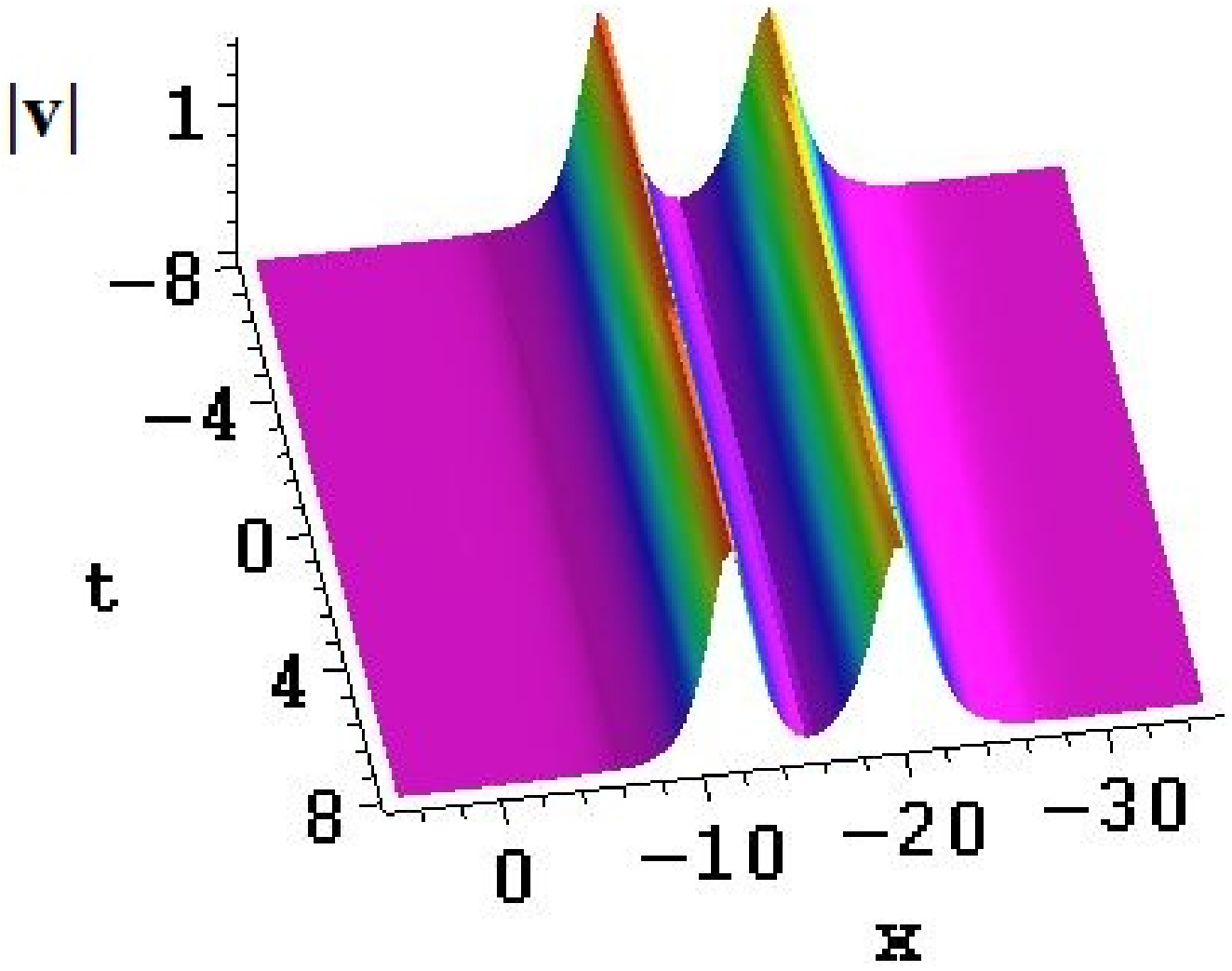}}
\begin{center}
\hskip 1cm $(\rm{a})$ \hskip 5cm $(\rm{b})$
\end{center}
\caption{Evolution plot of the second-order dark-bright-rogue wave in coupled Hirota equations with the parameters chosen by
$\epsilon=1/100,\alpha=1/10000,d_{1}=1,d_{2}=0,m_{1}=0,n_{1}=0$. (a) two dark solitons and a fundamental second-order rogue wave separate
in $u$ component; (b) two bright solitons and a fundamental second-order rogue wave separate
in $v$ component. The rogue wave in $v$ component is difficult to be seen for its zero-amplitude background wave.}
\centering
\renewcommand{\figurename}{{\bf Fig.}}
{\includegraphics[height=4cm,width=6cm]{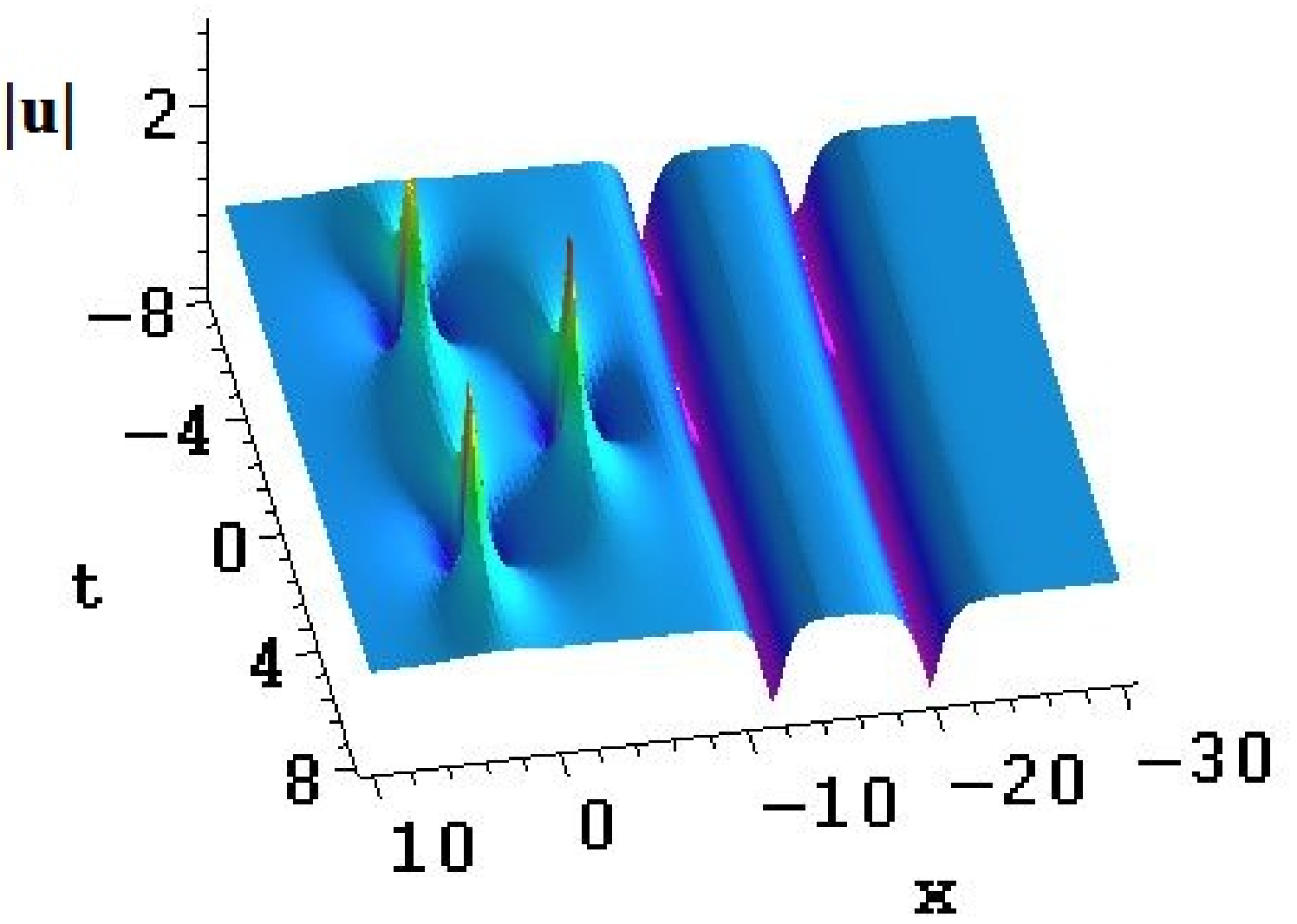}}
{\includegraphics[height=4cm,width=6cm]{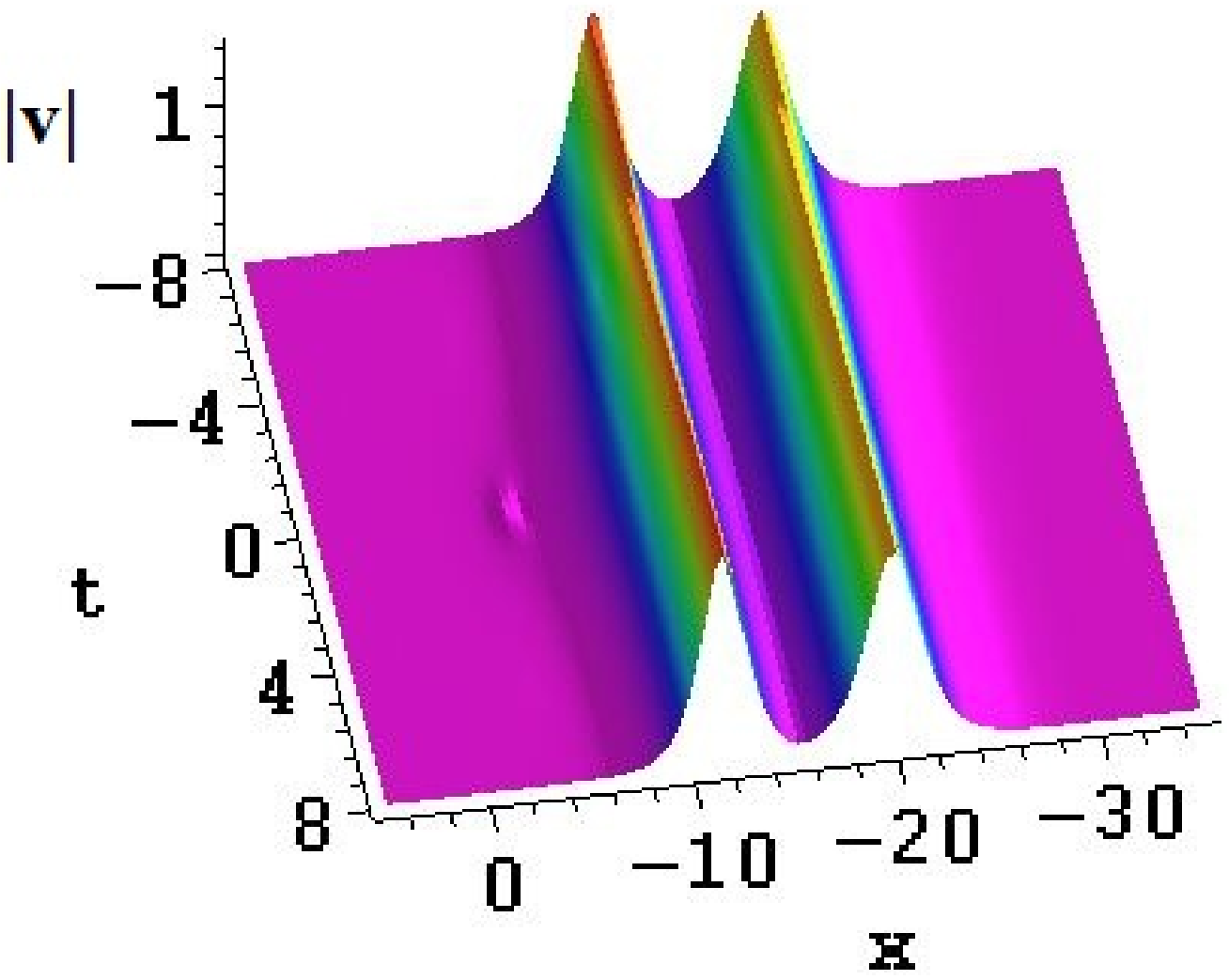}}
\begin{center}
\hskip 1cm $(\rm{a})$ \hskip 5cm $(\rm{b})$
\end{center}
\caption{Evolution plot of the second-order dark-bright-rogue wave in coupled Hirota equations with the parameters chosen by
$\epsilon=1/100,\alpha=1/10000,d_{1}=1,d_{2}=0,m_{1}=10,n_{1}=0$. (a) two dark solitons together with
a second-order rogue wave of triangular pattern in $u$ component;
(b) two bright solitons together with a second-order rogue wave of triangular pattern
in $v$ component.  The rogue waves in $v$ component
are difficult to be seen for its zero-amplitude background wave.}
\end{figure}

\begin{figure}
\centering
\renewcommand{\figurename}{{\bf Fig.}}
{\includegraphics[height=4cm,width=6cm]{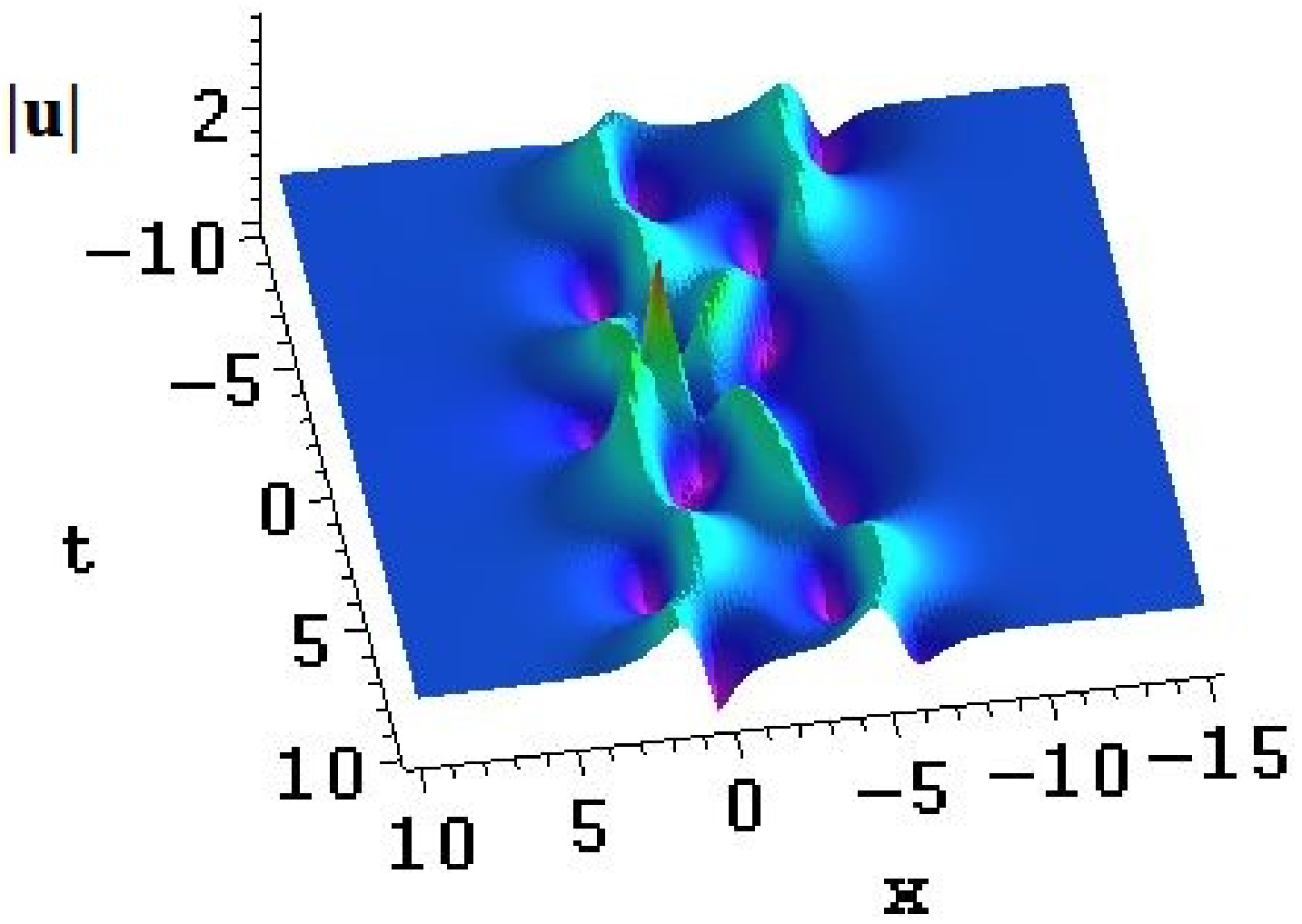}}
{\includegraphics[height=4cm,width=6cm]{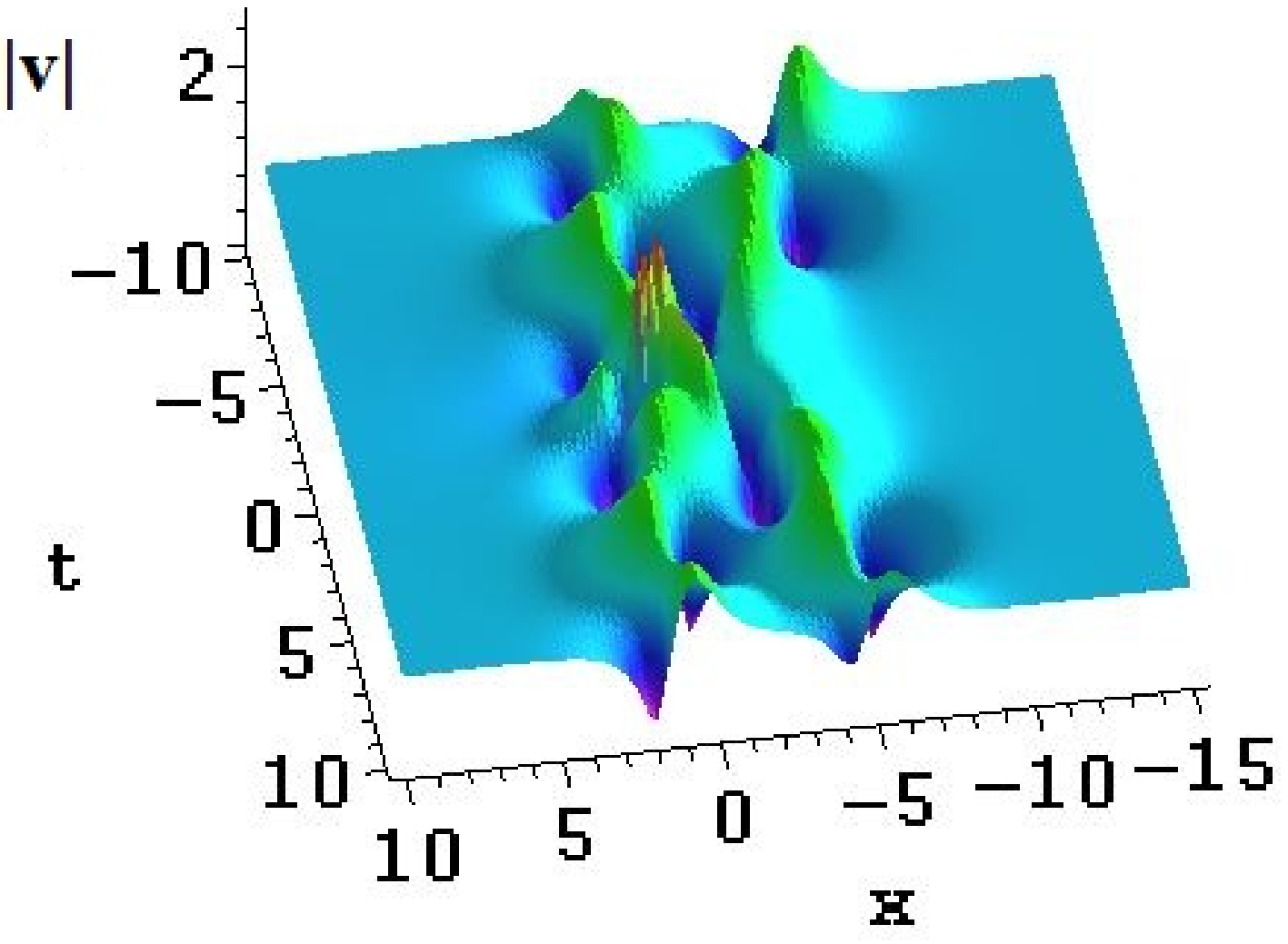}}
\begin{center}
\hskip 1cm $(\rm{a})$ \hskip 5cm $(\rm{b})$
\end{center}
\caption{Evolution plot of the second-order breather-rogue wave in coupled Hirota equations with the parameters chosen by
$\epsilon=1/100,\alpha=10,d_{1}=1,d_{2}=1,m_{1}=0,n_{1}=0$.
(a) two Akhmediev breathers merge with a fundamental second-order rogue wave in $u$ component;
(b) two Akhmediev breathers merge with a fundamental second-order rogue wave in $v$ component.}
\centering
\renewcommand{\figurename}{{\bf Fig.}}
{\includegraphics[height=4cm,width=6cm]{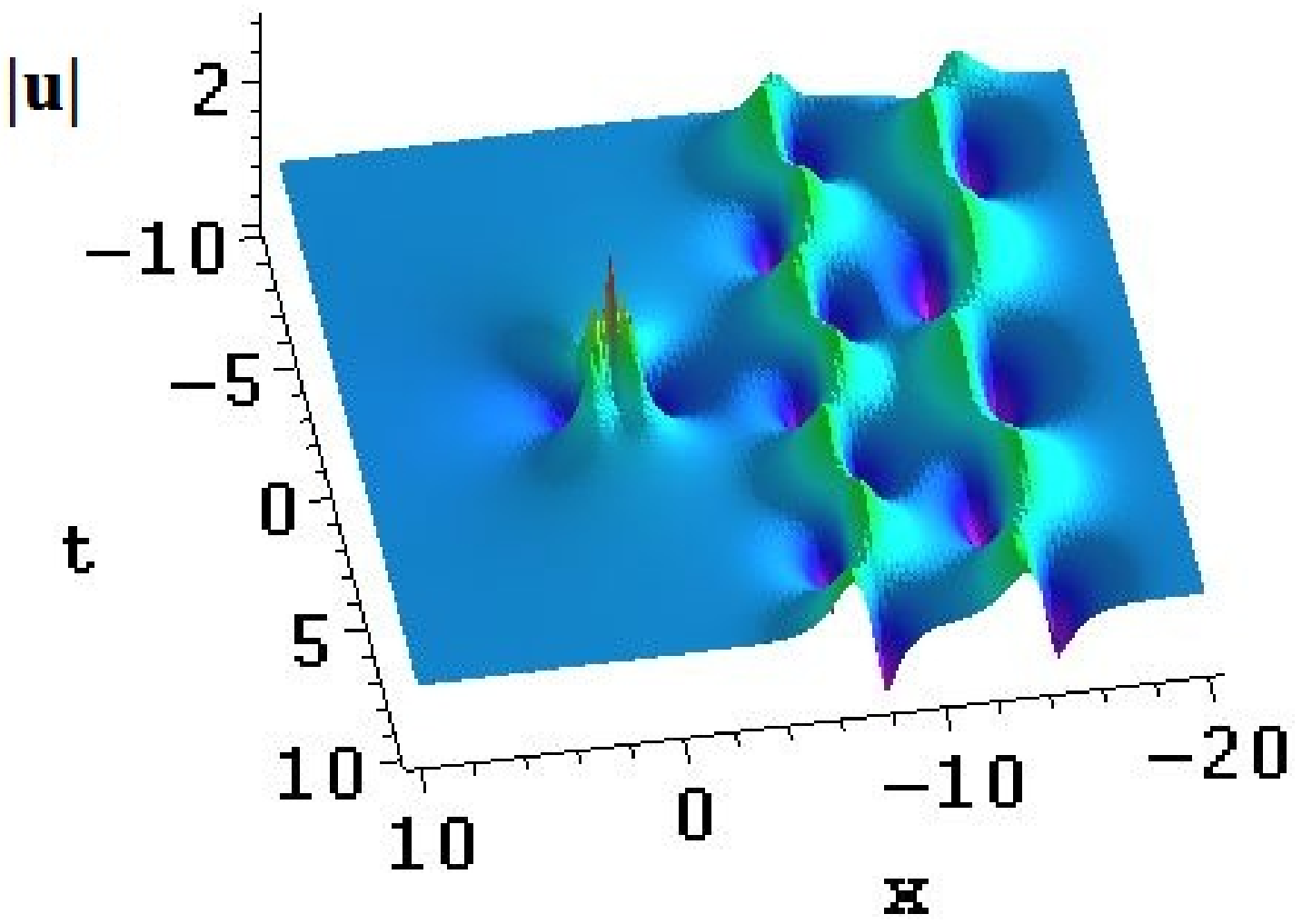}}
{\includegraphics[height=4cm,width=6cm]{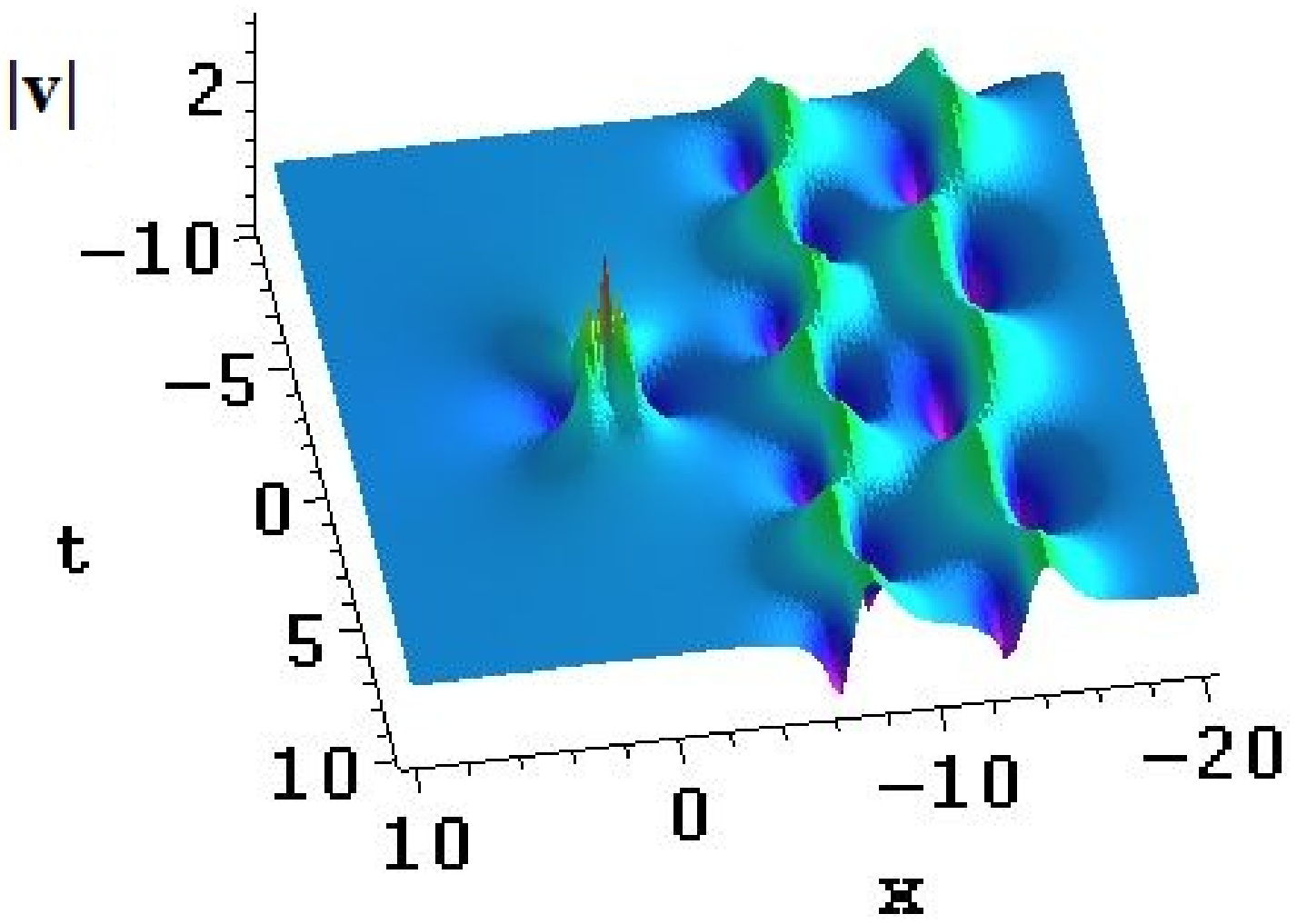}}
\begin{center}
\hskip 1cm $(\rm{a})$ \hskip 5cm $(\rm{b})$
\end{center}
\caption{Evolution plot of the second-order breather-rogue wave in coupled Hirota equations with the parameters chosen by
$\epsilon=1/100,\alpha=1/10000,d_{1}=1,d_{2}=1,m_{1}=0,n_{1}=0$.
(a) two parallel Akhmediev breathers and a fundamental second-order rogue wave separate in $u$ component;
(b) two parallel Akhmediev breathers and a fundamental second-order rogue wave separate in $v$ component.}
\centering
\renewcommand{\figurename}{{\bf Fig.}}
{\includegraphics[height=4cm,width=6cm]{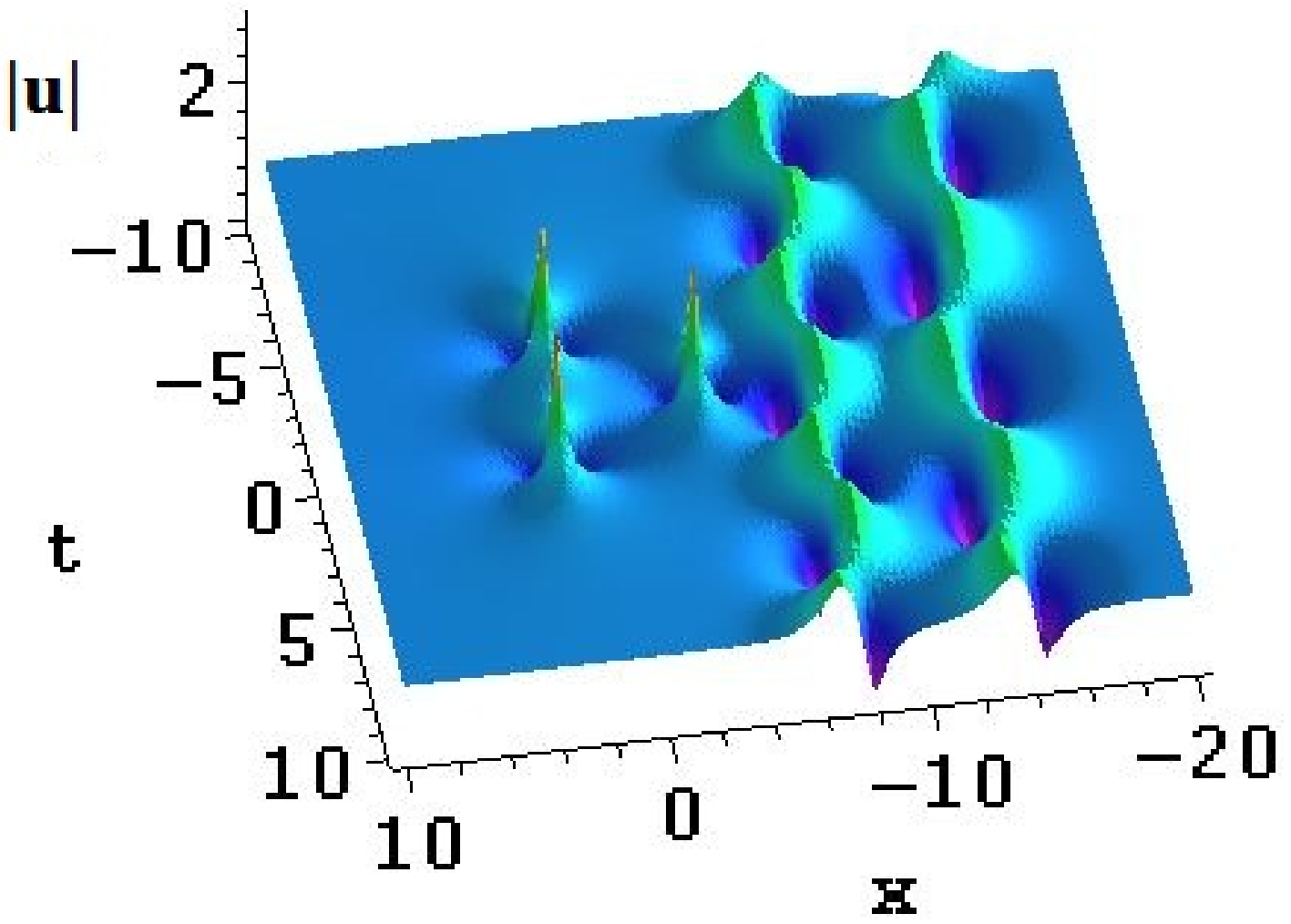}}
{\includegraphics[height=4cm,width=6cm]{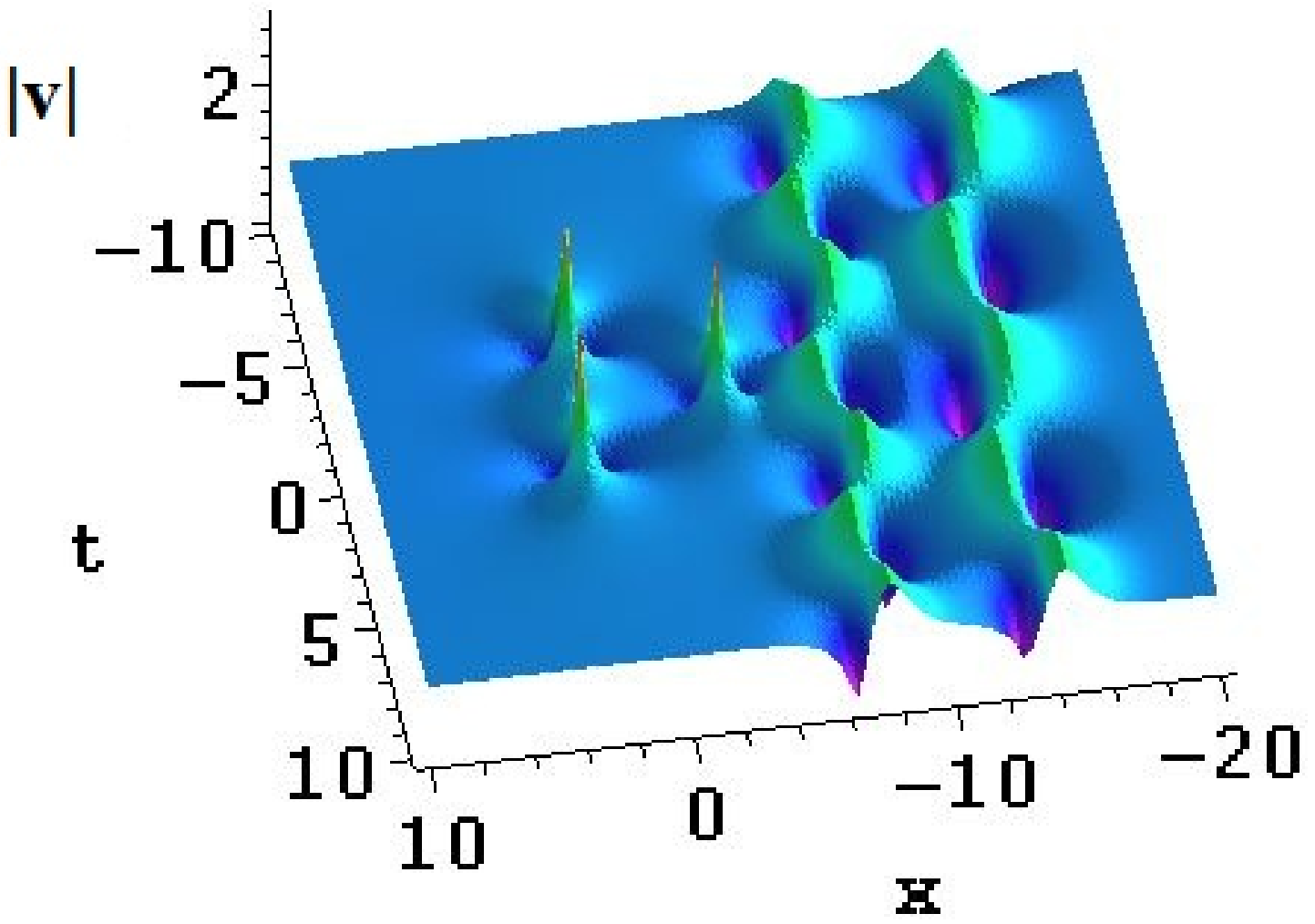}}
\begin{center}
\hskip 1cm $(\rm{a})$ \hskip 5cm $(\rm{b})$
\end{center}
\caption{Evolution plot of the second-order breather-rogue wave in coupled Hirota equations with the parameters chosen by
$\epsilon=1/100,\alpha=1/10000,d_{1}=1,d_{2}=1,m_{1}=10,n_{1}=0$.
(a) two parallel Akhmediev breathers and a second-order rogue wave of triangular pattern separate in $u$ component;
(b) two parallel Akhmediev breathers and a second-order rogue wave of triangular pattern separate in $v$ component.}
\end{figure}

\newpage
\section*{Acknowledgment}
The project is supported by the National Natural Science Foundation of China (Grant No.
11275072), Research Fund for the Doctoral Program of Higher Education of China (No. 20120076110024),
the Innovative Research Team Program of the National Natural Science Foundation of China (Grant No. 61321064),
Talent Fund and K.C. Wong Magna Fund in Ningbo University.


\section*{References}
\begin{thebibliography}{99}
\itemsep=-1pt plus.1pt minus.1pt
\small

\bibitem{ad01}L. Khaykovich, F. Schreck, G. Ferrari, T. Bourdel, J. Cubizolles,
L. D. Carr, Y. Castin, C. Salomon, Formation of a matter-wave bright soliton, Science 296 (2002) 1290-1293.
\bibitem{ad02}S. Burger, K. Bongs, S. Dettmer, W. Ertmer, K. Sengstock, Dark solitons in Bose-Einstein condensates, Phys. Rev. Lett. 83 (1999) 5198-5201.
\bibitem{ad03}N. Akhmediev, V.M. Eleonskii, N.E. Kulagin, Exact first-order solutions of the nonlinear Schr\"{o}dinger equation, Theor. Math. Phys. 72 (1987) 183-196.
\bibitem{ad04}Y.C. Ma, The perturbed plane-wave solutions of the cubic Schr\"{o}dinger equation, Stud. Appl. Math. 60 (1979) 43-58.
\bibitem{az01}K.W. Chow, R.H.J. Grimshaw, E. Ding, Interactions of breathers and solitons in the extended Korteweg-de Vries equation, Wave Motion 43 (2005) 158-166.
\bibitem{w1}V.S. Gerdjikov, R.I. Ivanov, A.V. Kyuldjiev, On the N-wave equations and soliton interactions in two and three dimensions,
            Wave Motion 48 (2011) 791-804.
\bibitem{w2}A.V. Porubov, H. Tsuji, I.V. Lavrenov, M. Oikawa, Formation of the rogue wave due to non-linear two-dimensional waves interaction, Wave Motion 42 (2005) 202-210.
\bibitem{w3}C. Fochesato, S. Grilli, F. Dias, Numerical modeling of extreme rogue waves generated by directional energy focusing,
       Wave Motion 44 (2007) 395-416.
\bibitem{1}C. Garrett, J. Gemmrich, Rogue waves, Phys. Today 62 (2009) 62-63. 
\bibitem{2}D.R. Solli, C. Ropers, P. Koonath, B. Jalali, Optical rogue waves, Nature 450 (2007) 1054-1057. 
\bibitem{3}N. Akhmediev, J.M. Dudley, D.R Solli, S.K. Turitsyn, Recent progress in investigating optical rogue waves, J. Opt. 15 (2013) 060201. 
\bibitem{6}N. Akhmediev, A. Ankiewicz, M. Taki, Waves that appear from nowhere and disappear without a trace,
            Phys. Lett. A 373 (2009) 675-678. 
\bibitem{4}D.H. Peregrine, J. Austral, Water waves, nonlinear Schr\"{o}dinger equations and their solutions, Math. Soc. B:  Appl. Math. 25 (1983) 16-43. 
\bibitem{5}N. Akhmediev, A. Ankiewicz,  J.M. Soto-Crespo,  Rogue waves and rational solutions of the nonlinear
              Schr\"{o}dinger equation, Phys. Rev. E 80 (2009) 026601.  
\bibitem{7}A. Ankiewicz, D.J. Kedziora, N. Akhmediev, Rogue wave triplets, Phys. Lett. A 373 (2011) 2782-2785.
\bibitem{8}D.J. Kedziora, A. Ankiewicz,  N. Akhmediev, Circular rogue wave clusters, Phys. Rev. E 84 (2011) 056611. 
\bibitem{9}D.J. Kedziora, A. Ankiewicz,  N. Akhmediev, Classifying the hierarchy of nonlinear-Schr\"{o}dinger-equation rogue-wave solutions, Phys. Rev. E 88 (2013) 013207. 
\bibitem{10}D.J. Kedziora, A. Ankiewicz,  N. Akhmediev, Second-order nonlinear Schr\"{o}dinger equation breather solutions in the degenerate and rogue wave limits, Phys. Rev. E 85 (2012) 066601. 
\bibitem{11}A. Chabchoub, N.P. Hoffmann, N. Akhmediev, Rogue wave observation in a water wave tank, Phys. Rev. Lett. 106 (2011) 204502. 
\bibitem{21} P. Gaillard,  Families of quasi-rational solutions of the NLS equation and multi-rogue waves, J. Phys. A  44 (2011) 435204.
\bibitem{22}B.L. Guo, L.M. Ling, Q.P. Liu, Nonlinear Schr\"{o}dinger equation: Generalized Darboux transformation and rogue wave solutions, Phys. Rev. E 85 (2012) 026607. 
\bibitem{18}S.W. Xu, J.S. He,  The Darboux transformation of the derivative nonlinear Schr\"{o}dinger equation,
J. Phys. A  44 (2011) 305203. 
\bibitem{23}B.L. Guo, L.M. Ling, Q.P. Liu, High-Order solutions and generalized Darboux transformations of derivative nonlinear Schr\"{o}dinger Equations, Stud. Appl. Math. 130  (2013) 317-344. 
\bibitem{19}Y.S. Tao, J.S. He, Multisolitons, breathers, and rogue waves for the Hirota equation generated by the Darboux transformation, Phys. Rev. E 85 (2012) 026601.
\bibitem{32}U. Bandelow,  N. Akhmediev,  Persistence of rogue waves in extended nonlinear Schr\"{o}dinger equations: integrable Sasa-Satsuma case, Phys. Lett. A 376 (2012) 1558-1561.  
\bibitem{31}Z.Y. Yan,  Nonautonomous \lq\lq rogons\rq\rq in the inhomogeneous nonlinear Schr\"{o}dinger equation with variable coefficients, Phys. Lett. A 374 (2010) 672-679. 
\bibitem{30}A. Ankiewicz,  N. Akhmediev,  J.M. Soto-Crespo,  Discrete rogue waves of the Ablowitz-Ladik and Hirota equations, Phys. Rev. E  82 (2010) 026602.  
\bibitem{37}Y. Ohta, J.K. Yang, Dynamics of rogue waves in the Davey-Stewartson II equation, J. Phys. A 46 (2013) 105202. 
\bibitem{27}F. Baronio,  A. Degasperis, M. Conforti, S. Wabnitz, Solutions of the vector nonlinear Schr\"{o}dinger equations: evidence for deterministic rogue waves, Phys. Rev. Lett. 109 (2012) 044102. 
\bibitem{29}L.C. Zhao,  J. Liu,  Rogue-wave solutions of a three-component coupled nonlinear Schr\"{o}dinger equation, Phys. Rev. E   87 (2013) 013201.  
\bibitem{exx01}B.G. Zhai, W.G. Zhang, X.L. Wang, H.Q. Zhang, Multi-rogue waves and rational solutions of the coupled nonlinear Schr\"{o}dinger equations, Nonlinear Anal. RWA 14 (2013) 14-27.
\bibitem{26}Y.V. Bludov, V.V. Konotop, N. Akhmediev, Vector rogue waves in binary mixtures of Bose-Einstein condensates, Eur. Phys. J. Spec. Top. 185 (2010) 169-180. 
\bibitem{24}B.L. Guo, L.M. Ling, Rogue wave, breathers and bright-dark-rogue solutions for the coupled Schr\"{o}dinger equations, Chin. Phys. Lett. 28 (2011) 110202. 
\bibitem{ex02}L.C. Zhao, J. Liu, Localized nonlinear waves in a two-mode nonlinear fiber,
J. Opt. Soc. Am. B 29 (2012) 3119-3127.
\bibitem{mu}G. Mu, Z.Y. Qin, R. Grimshaw, Dynamics of rogue waves on a multi-soliton background in a vector nonlinear Schr\"{o}dinger equation, (2014) arXiv:1404.2988.  
\bibitem{16}C.H. Gu, H.S. Hu,  Z.X. Zhou, Darboux transformations in integrable systems: theory and their applications to geometry, Springer-Verlag, New York, 2005.
\bibitem{17}V.B. Matveev, M.A. Salle, Darboux transformations and solitons, Springer-Verlag, Berlin-Heidelberg, 1991.
\bibitem{dt1}Y.S. Li, W.X. Ma, J.E. Zhang, Darboux transformations of classical Boussinesq system and its multi-soliton solutions,
             Phys. Lett. A  284 (2001) 253-258.
\bibitem{dt2}E.G. Fan, Darboux transformation and soliton-like solutions for the Gerdjikov-Ivanov equation, J. Phys. A
              33 (2000) 6925.
\bibitem{dt3}B. Xue, X. Wang, The Darboux transformation and new explicit solutions for the Belov-Chaltikian
             lattice, Chin. Phys. Lett. 29 (2012) 100201.
\bibitem{dt4}X. Wang, Y. Chen, Darboux transformations and N-soliton solutions of two (2+1)-Dimensional
             nonlinear equations, Commum. Theor. Phys. 61 (2014) 423-430.
\bibitem{AB} X. Wang, Y.Q. Li, Y. Chen, Rogue wave solutions in AB system, (2013) arXiv:1312.6479.
\bibitem{ex01}Zhaqilao, Nth-order rogue wave solutions of the complex modified Korteweg-de Vries equation,
Phys. Scr.  87 (2013) 065401.
\bibitem{33}L.M. Ling, B.L. Guo, L.C. Zhao, High-order Rogue Waves in Vector Nonlinear Schr\"{o}dinger Equations,
Phys. Rev. E 89 (2014) 041201.
\bibitem{39}R.S. Tasgal, M. J. Potasek, Soliton solutions to coupled higher-order nonlinear Schr\"{o}dinger equations, J. Math. Phys. 33 (1992) 1208-1215. 
\bibitem{41} S.G. Bindu, A. Mahalingam, K. Porsezian, Dark soliton solutions of the coupled Hirota equation in nonlinear fiber, Phys. Lett. A 286 (2001) 321-331. 
\bibitem{40}K. Porsezian, K. Nakkeeran, Optical solitons in birefringent fibre-Backlund transformation approach,
Pure Appl. Opt. 6 (1997) L7-L11. 
\bibitem{28}S.H. Chen, L.Y. Song,  Rogue waves in coupled Hirota systems, Phys. Rev. E  87 (2013) 032910.  
\bibitem{co}D.J. Kedziora, A. Ankiewicz, N. Akhmediev, Rogue waves and solitons on a cnoidal background, Eur. Phys. J-Spec. Top
223 (2014) 43-62.
\end{thebibliography}
\end{document}